\documentclass[pra,aps,twocolumn,nopacs,superscriptaddress,nofootinbib]{revtex4}

\usepackage{graphicx}
\usepackage{dcolumn}
\usepackage{bm}
\usepackage{bbm}
\usepackage{amsmath}
\usepackage{epsfig}
\usepackage{indentfirst}
\usepackage{psfrag}
\usepackage{subfigure}
\usepackage{amssymb}
\usepackage{color}
\usepackage[colorlinks,linkcolor=blue,citecolor=blue,urlcolor=blue,hyperindex,driverfallback=dvipdfm]{hyperref}
\usepackage[T1]{fontenc}
\usepackage{comment}
\usepackage{upgreek}
\usepackage{soul}

\def\ii{{\rm i}}  \def\ee{{\rm e}}
  
\def\rb{{\bf r}}  \def\Rb{{\bf R}}    \def\vb{{\bf v}}
\def\xx{\hat{\bf x}}  \def\yy{\hat{\bf y}}  \def\zz{\hat{\bf z}}  
  \def\eh{\hat{\bf e}}
\def\RR{\hat{\bf R}}  
\def\kb{{\bf k}}  \def\kpar{k_\parallel}  \def\kparb{{\bf k}_\parallel}
\def\Qb{{\bf Q}}    
\def\me{m_{\rm e}}  
\def\Eb{{\bf E}}      \def\Ab{{\bf A}}

\def\kperp{k_\perp}  \def\kperpb{{\bf k}_\perp}

\begin{document}
\title{Optical Modulation of Electron Beams in Free Space}
\author{F.~Javier~Garc\'{\i}a~de~Abajo}
\email[Corresponding author: ]{\\ javier.garciadeabajo@nanophotonics.es}
\affiliation{ICFO-Institut de Ciencies Fotoniques, The Barcelona Institute of Science and Technology, 08860 Castelldefels (Barcelona), Spain}
\affiliation{ICREA-Instituci\'o Catalana de Recerca i Estudis Avan\c{c}ats, Passeig Llu\'{\i}s Companys 23, 08010 Barcelona, Spain}
\author{Andrea~Kone\v{c}n\'{a}}
\affiliation{ICFO-Institut de Ciencies Fotoniques, The Barcelona Institute of Science and Technology, 08860 Castelldefels (Barcelona), Spain}

\begin{abstract}
We exploit free-space interactions between electron beams and tailored light fields to imprint on-demand phase profiles on the electron wave functions. Through rigorous semiclassical theory involving a quantum description of the electrons, we show that monochromatic optical fields focused in vacuum can be used to correct electron beam aberrations and produce selected focal shapes. Stimulated elastic Compton scattering is exploited to imprint the phase, which is proportional to the integrated optical field intensity along the electron path and depends on the transverse beam position. The required light intensities are attainable in currently available ultrafast electron microscope setups, thus opening the field of free-space optical manipulation of electron beams.
\end{abstract}
\date{\today}
\maketitle


\section{Introduction}

Electron microscopy has experienced impressive advances over the last decades thanks to the design of sophisticated magnetostatic and electrostatic lenses that reduce electron optics aberrations \cite{HRU98,BDK02,HS19} and are capable of focusing electron beams (e-beams) with sub{\aa}ngstrom accuracy \cite{NCD04,MKM08}. In addition, the availability of more selective monochromators \cite{KLD14} enable the exploration of sample excitations down to the mid-infrared regime \cite{LTH17,HNY18,HKR19,HHP19}. Such precise control over e-beam shape and energy is crucial for atomic-scale imaging and spectroscopy \cite{HRU98,BDK02,HS19,KLD14,NCD04,MKM08,LTH17,HNY18,HKR19,HHP19}.

The focused e-beam profile ultimately depends on the phase acquired by the electron along its passage through the microscope column. Besides electron optics lenses, several physical elements have been demonstrated to control transverse e-beam shaping. In particular, biprisms based on biased wires provide a dramatic example of laterally-varying phase imprinting that is commonly used for e-beam splitting in electron holography \cite{MD1956}, along with applications such as symmetry-selected plasmon excitation in metallic nanowires \cite{GBL17}. In a related context, vortex e-beams have been realized using a magnetic pseudo-monopole \cite{BVV14}. Recently, plates with individually-biased perforations have been developed to enable position-selective control over the electric Aharonov-Bohm phase stamped on the electron wave function \cite{VBM18}, while passive carved plates have been employed as amplitude filters to produce highly-chiral electron vortices \cite{VTS10,MAA11,SLL14} and aberration correctors \cite{GTY17,SRP18}.

The electron phase can also be modified by the ponderomotive force associated with the interaction between e-beams and optical fields. In particular, periodic light standing waves were predicted to produce electron diffraction \cite{KD1933}, eventually observed in a challenging experiment \cite{FAB01,FB02,B07} that had to circumvent the weak free-space electron-photon coupling associated with energy-momentum mismatch \cite{paper149}. Such mismatch forbids single photon emission or absorption processes by free electrons, consequently limiting electron-light coupling to second-order interactions that concatenate an even number of virtual photon events. This type of interaction has been recently exploited to produce attosecond free-electron pulses \cite{KSH18,K19_2}. Interestingly, the presence of material structures introduces scattered optical fields that can supply momentum and break the mismatch, thus enabling real photon processes \cite{paper149}, used for example in laser-driven electron accelerators \cite{BNM19,SMY19}. Because the strength of scattered fields reflects the nanoscale optical response of the materials involved, this was speculated to enable electron energy-gain spectroscopy as a way to dramatically improve spectral resolution in electron microscopes \cite{H99,paper114,H09}, as later corroborated in experiment \cite{paper306}. By synchronizing the arrival of electron and light pulses at the sample, photon-induced near-field electron microscopy (PINEM) was demonstrated to exert temporal control over the electron wave function along the beam direction \cite{BFZ09,paper151,PLZ10,PZ12,KGK14,PLQ15,FES15,paper282,EFS16,RB16,VFZ16,KML17,FBR17,PRY17,paper311,MB18,MB18_2,paper325,paper332,K19,paper339,T20,KLS20,WDS20}. Additionally, modulation of the transverse wave function can be achieved in PINEM by laterally shaping the employed light \cite{paper272}, which results in the transfer of linear \cite{paper311} and angular \cite{paper332,paper312} momentum between photons and electrons.

Recently, we have proposed to use PINEM to imprint on-demand transverse e-beam phase profiles \cite{paper351}, thus relying on ultrafast e-beam shaping as an alternative approach to aberration correction. This method enables fast active control over the modulated e-beam at the expense of retaining only $\sim1/3$ of monochromatic electrons and potentially introducing decoherence through inelastic interactions with the light scatterer. An approach to phase moulding in which no materials are involved and the electron energy is preserved would then be desirable.

In this Letter, we propose an optical free-space electron modulator (OFEM) in which a phase profile is imprinted on the transverse electron wave function by means of stimulated elastic Compton scattering associated with the $A^2$ term in the light-electron coupling Hamiltonian. The absence of material structures prevents electron decoherence and enables the use of high light intensities, as required to activate ponderomotive forces. We present a simple, yet rigorous semiclassical theory that supports applications in aberration correction and transverse e-beam shaping. While optical e-beam phase stamping has been demonstrated using a continuous-wave laser in a tour-de-force experiment \cite{SAC19}, we envision pulsed illumination as a more feasible route to implement an OFEM, exploiting recent advances in ultrafast electron microscopy, particularly in systems that incorporate light injection with high numerical aperture \cite{paper325} for diffraction-limited patterning of the optical field.

\section{Free-space optical phase imprinting}

We study the free-space interaction between an e-beam and a light field represented by its vector potential $\Ab(\rb,t)$. Taking the e-beam propagation direction along $z$, it is convenient to write the electron wave function as $\psi_z(\Rb,t)=\ee^{\ii q_0z-\ii E_0t/\hbar}\phi(\rb,t)$, where $\Rb=(x,y)$ denotes transverse coordinates and we separate the slowly-varying envelope $\phi(\rb,t)$ from the fast oscillations imposed by the central wave vector $q_0$ and energy $E_0$. Under typical conditions met in electron microscopes, and assuming that interaction with light only produces small variations in the electron energy compared with $E_0$, we can adopt the nonrecoil approximation to reduce the Dirac equation in the minimal coupling scheme to an effective Schr\"odinger equation \cite{ValerioCompton},
\begin{align}
\left(\partial_t+\vb\cdot\nabla\right)\phi(\rb,t)=\frac{-\ii}{\hbar}\mathcal{H}'(\rb,t)\phi(\rb,t),
\nonumber
\end{align}
where
\begin{align}
\mathcal{H}'=\frac{e}{c}\vb\cdot\Ab+\frac{e^2}{2\me c^2\gamma}\left(A_x^2+A_y^2+\frac{1}{\gamma^2}A_z^2\right)
\nonumber
\end{align}
is the interaction Hamiltonian, $\vb=v\zz$ is the electron velocity, and $\gamma=1/\sqrt{1-v^2/c^2}$ introduces relativistic corrections to the $A^2$ term. This equation admits the analytical solution
\begin{align}
\phi(\rb,t)=\phi_0(\rb-\vb t)\exp\left[\frac{-\ii}{\hbar}\int_{-\infty}^t \!\!\! dt' \;\mathcal{H}'(\rb-\vb t+\vb t',t')\right],
\nonumber
\end{align}
where $\phi_0(\rb-\vb t)$ is the incident electron wave function.

We consider that the light field acts on the electron over a sufficiently short path length as to neglect any transverse variation in its wave function during interaction. We also note that the $\vb\cdot\Ab$ term in $\mathcal{H}'$ does not contribute to the final electron state because it represents real photon absorption or emission events, which are kinematically forbidden. Likewise, following a similar argument, under monochromatic illumination with light of frequency $\omega$, the time-varying components in $A^2$ ($\propto\ee^{\pm2\ii\omega t}$), which describe two-photon emission or absorption, also produce vanishing contributions. The remaining terms $\propto\ee^{\pm\ii\omega t\mp\ii\omega t}$ represent stimulated elastic Compton scattering, a second-order process that combines virtual absorption and emission of photons, amplified by the large population of their initial and final states. An alternative description of this effect is provided by the ponderomotive force acting on a classical point-charge electron and giving rise to diffraction in the resulting effective potential \cite{B07}. As we are interested in imprinting a phase on the electron wave function without altering its energy, we consider spectrally narrow external illumination that can be effectively regarded as monochromatic, such as that produced by laser pulses of much longer duration than the electron pulse. Writing the external field as $\Eb(\rb,t)=2{\rm Re}\{\Eb(\rb)\ee^{-\ii\omega t}\}$, the transmitted wave function becomes
\begin{align}
\psi_z(\Rb,t)=\psi_0(\rb-\vb t)\;\ee^{\ii\varphi(\Rb)},
\nonumber
\end{align}
where
\begin{align}
\varphi(\Rb)\!=&\frac{-1}{\mathcal{M}\omega^2} \!\!\!\int_{-\infty}^\infty \!\!\!\!\!dz \bigg[|E_x(\rb)|^2\!+\!|E_y(\rb)|^2\!+\!\frac{1}{\gamma^2}|E_z(\rb)|^2\bigg]
\label{phase}
\end{align}
is an imprinted phase that depends on transverse coordinates $\Rb$, we define the scaled mass $\mathcal{M}=\me\gamma v/c\alpha$, and $\alpha\approx1/137$ is the fine structure constant.

\begin{figure}
\centering{\includegraphics[width=0.5\textwidth]{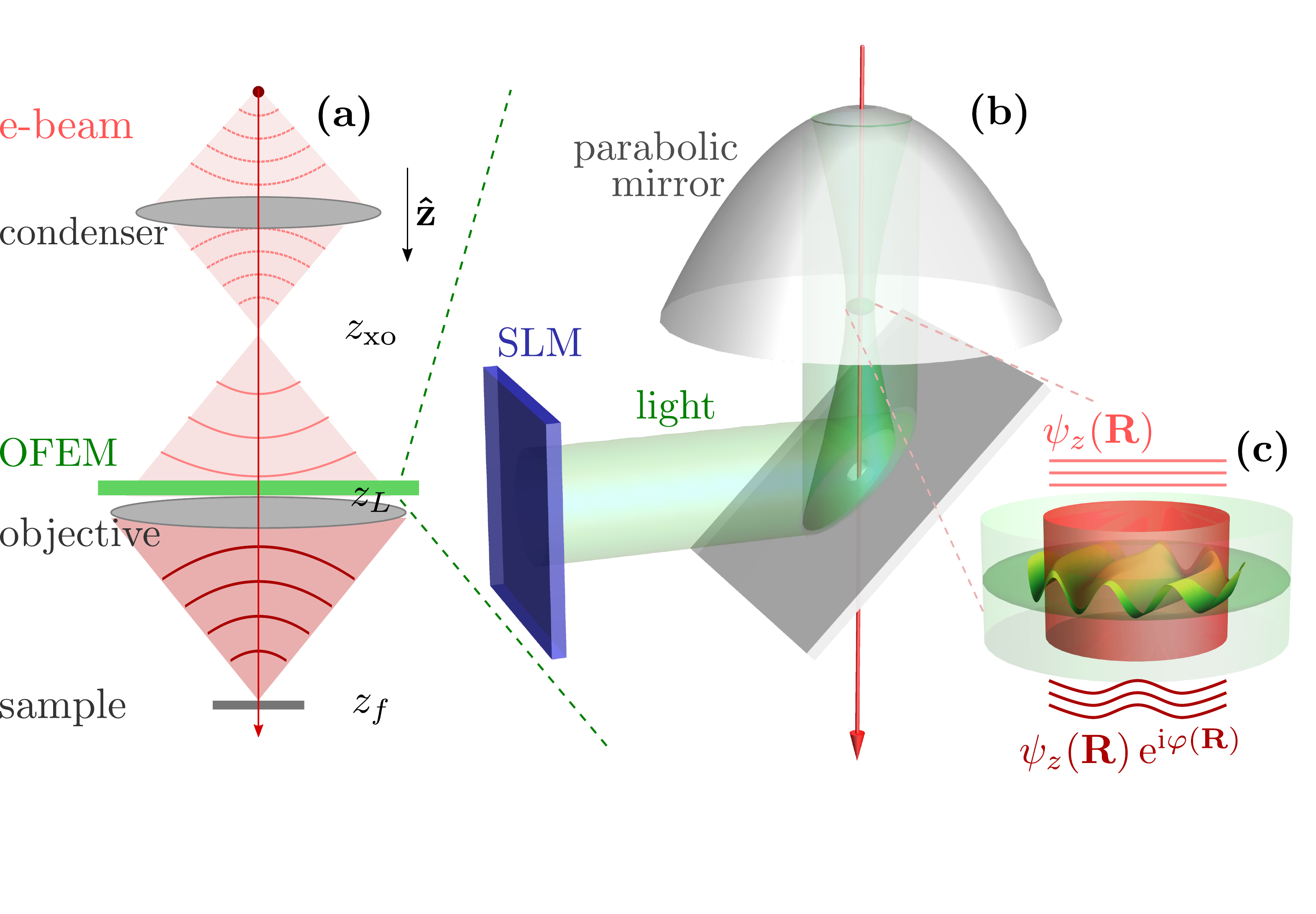}}
\caption{{\bf Optical free-space electron modulator (OFEM).} (a) The proposed element is placed in the electron microscope column right before the objective lens. (b) It incorporates a parabolic mirror that focuses light with high numerical aperture on a vacuum region that intersects the electron beam. The electric field distribution at the optical focal spot is patterned by using a far-field spatial light modulator (SLM). (c) A phase is imprinted on the electron wave function, whose dependence on transverse coordinates $\Rb$ is proportional to the field intensity integrated along $z$.}
\label{Fig1}
\end{figure}

\section{Description of an OFEM}

We envision an OFEM placed right before the objective lens of an electron microscope [Fig.\ \ref{Fig1}(a)], in a region where the e-beam spans a large diameter ($\gtrsim100$ times the light wavelength). The OFEM could consist of a combination of planar and parabolic mirrors with drilled holes that allow the e-beam to pass through the optical focal region [Fig.\ \ref{Fig1}(b)]. The electron is then exposed to intense fields that can be shaped with diffraction-limited lateral resolution through a spatial light modulator and a high numerical aperture of the parabolic mirror. This results in a controlled position-dependent phase, as prescribed by Eq.\ (\ref{phase}) [Fig.\ \ref{Fig1}(c)]. Considering a monochromatic e-beam and omitting for simplicity an overall $\ee^{-\ii E_0t/\hbar}$ time-dependent factor, free propagation of the electron wave function between planes $z$ and $z_f$ is described by
\begin{align}
\psi_{z_f}(\Rb_f)&=\iint\frac{d^2\Qb\,d^2\Rb}{(2\pi)^2}\,\ee^{\ii\Qb\cdot(\Rb_f-\Rb)+\ii q_z(z_f-z)}\psi_{z}(\Rb) \nonumber\\
&\propto\int d^2\Rb\;\ee^{\ii q_0|\Rb_f-\Rb|^2/2(z_f-z)}\psi_{z}(\Rb),
\label{psi1}
\end{align}
where the second line is obtained by performing the $\Qb=(q_x,q_y)$ integral in the paraxial approximation (i.e., $q_z=\sqrt{q_0^2-Q^2}\approx q_0-Q^2/2q_0$) and we are interested in exploring positions $\rb_f=(\Rb_f,z)$ near an electron focal point. In a simplified microscope model, we take $z_L$ at the entrance of the objective lens where the OFEM is also placed, and the incident electron is a spherical wave $\psi_{z_L}(\Rb)\propto\ee^{\ii q_0R^2/2(z_L-z_{\rm xo})}$ emanating from the crossover point $\rb=(0,0,z_{\rm xo})$ following the condenser lens. Introducing in Eq.\ (\ref{psi1}) the phase $\ee^{-\ii q_0 R^2/2f}$ produced by an objective lens with focal distance $f$ and aperture radius $R_{\rm max}$, we have (see Appendix\ \ref{secS1})
\begin{align}
\psi_{z_f}(\Rb_f)&\propto\int\!d^2\vec{\theta}\,\ee^{-\ii q_0\vec{\theta}\cdot\Rb_f}\ee^{\ii\chi(\vec{\theta})+\ii\varphi(\Rb)}\ee^{\ii q_0R^2\Delta/2},
\label{psifinal}
\end{align}
where $\vec{\theta}=\Rb/(z_f-z_L)$, we define $\Delta=1/(z_f-z_L)+1/(z_L-z_{\rm xo})-1/f$, the phases $\chi$ and $\varphi$ are produced by aberrations and the OFEM [Eq.\ (\ref{phase})], respectively, and the integral is restricted to $\theta<R_{\rm max}/(z_f-z_L)$. In what follows, we study the electron wave function profile $\psi_{z_f}(\Rb_f)$ as given by Eq.\ (\ref{psifinal}) at the focal plane $z_f$, defined by the condition $\Delta=0$.

{\it Required light intensity.---}From Eq.\ (\ref{phase}), the imprinted phase shift scales with the light intensity $I_0=c\,|E|^2/2\pi$ roughly as $\varphi/I_0\sim-2\pi L/\mathcal{M}c\omega^2$, where $L$ is the effective length of light-electron interaction, which depends on the focusing conditions of the external illumination. For example, for an electron moving along the axis of an optical paraxial vortex beam of azimuthal angular momentum number $m=1$ and wavelength $\lambda_0=2\pi c/\omega$, we have $L\approx2\lambda_0/\theta_L^2$, where $\theta_L$ is the light beam divergence half-angle (see Appendix\ \ref{secS3}). Under these conditions, a phase $\varphi=2\pi$ is achieved with a light power $\mathcal{P}=2\mathcal{M}c^2\omega\sim40\,$kW for 60\,keV electrons and $\lambda_0=500\,$nm; this result is independent of $\theta_L$, emphasizing the important role of phase accumulation along a long interaction region in a loosely focused light beam. This power level can be reached using nanosecond laser pulses synchronized with the electron passage through the optical field \cite{paper325}. We note that the phase scaling $\varphi\sim I_0/(v\gamma\omega^2)$ [see Eq.\ (\ref{phase})] leaves some room for improvement by placing the OFEM in low-energy regions of the e-beam to reduce the optical power demand.

\section{Aberration correction}

As an application of lateral phase moulding, we explore the conditions needed to compensate spherical aberration, corresponding to \cite{AOP01,PPB18} $\chi(\theta)=C_3q_0\theta^4/4$ in Eq.\ (\ref{psifinal}), where $C_3$ is a length coefficient. Upon examination of the phase profile imprinted by paraxial light vortex beams (see Appendix\ \ref{secS2}), we find that an azimuthal number $m=3$ produces the required radial dependence $\varphi(R)=-(\pi^2\,\mathcal{P}/96\,\mathcal{M}c^2\omega)\,(\theta_LR/\lambda_0)^4$ under the condition $R\ll\lambda_0/2\pi\theta_L$. For a typical microscope parameters $C_3=f=1\,$mm, 60\,keV electrons, $R_{\rm max}=30\,\upmu$m, and $\lambda_0=500\,$nm, the above condition is satisfied with $\theta_L\ll0.15^\circ$. Then, compensation of spherical aberration (i.e., $\varphi=-\chi$) is realized with a beam power $\mathcal{P}=(6\hbar c^2/\pi^4\alpha)\,C_3q_0^2\lambda_0^3/\theta_L^4(z_f-z_L)^4\sim3\times10^8\,$W, which is attainable using femtosecond laser pulses in ultrafast electron microscopes \cite{FES15,paper311,KLS20,WDS20}.

\section{Transverse e-beam shaping}

The production of on-demand electron spot profiles involves a two-step process comprising the determination of the necessary phase pattern $\varphi(\Rb)$, and from here the required optical beam parameters that generate that phase. While this is a complex task in general, we can find an approximate analytical solution for one-dimensional systems, assuming translational invariance along a direction $y$ perpendicular to the electron velocity. We therefore consider an optical beam characterized by an electric field $\Eb(\rb)=E(x,z)\yy$ and explore the generation of focal electron shapes defined by a wave function $\psi_z(x)$ independent of $y$. For light propagating along the positive $z$ direction, we can write without loss of generality $E(x,z)=\int_{-k_0}^{k_0}(dk_x/2\pi)\,\ee^{\ii(k_xx+k_zz)}\,\beta_{k_x}$ with $k_z=\sqrt{k_0-k_x^2}$ and $k_0=2\pi/\lambda_0$ in terms of the expansion coefficients $\beta_{k_x}$. Inserting this expression into Eq.\ (\ref{phase}), we find (see Appendix\ \ref{secS4}) $\varphi(x)=\varphi_0-(1/2\pi\mathcal{M}\omega^2)\int_{-k_0}^{k_0}dk_x\,\ee^{2\ii k_xx}(k_z/|k_x|)\,\beta_{k_x}\beta_{-k_x}^*$, where $\varphi_0$ is a global phase. Given a target profile $\varphi^{\rm target}(x)$, we can then use the approximation $\beta_{k_x}\beta_{-k_x}^*\approx-2\mathcal{M}\omega^2(|k_x|/k_z)\int dx\,\ee^{-2\ii k_xx}\,\varphi^{\rm target}(x)$ to generate the needed light beam coefficients. A particular solution is obtained by imposing $\beta_{k_x}=\beta_{-k_x}^*$, which renders $\beta_{k_x}$ as the square root of the right-hand side in the above expression. For any solution, combining these two integral expressions and dismissing $\varphi_0$, we find
\begin{align}
\varphi(x)=\frac{1}{\pi}\int_{-R_{\rm max}}^{R_{\rm max}}\!\!dx'\,\frac{\sin\left[4\pi(x-x')/\lambda_0\right]}{x-x'}\varphi^{\rm target}(x'),
\label{phifinal}
\end{align}
which yields a diffraction-limited phase profile.

\begin{figure}
\centering{\includegraphics[width=0.45\textwidth]{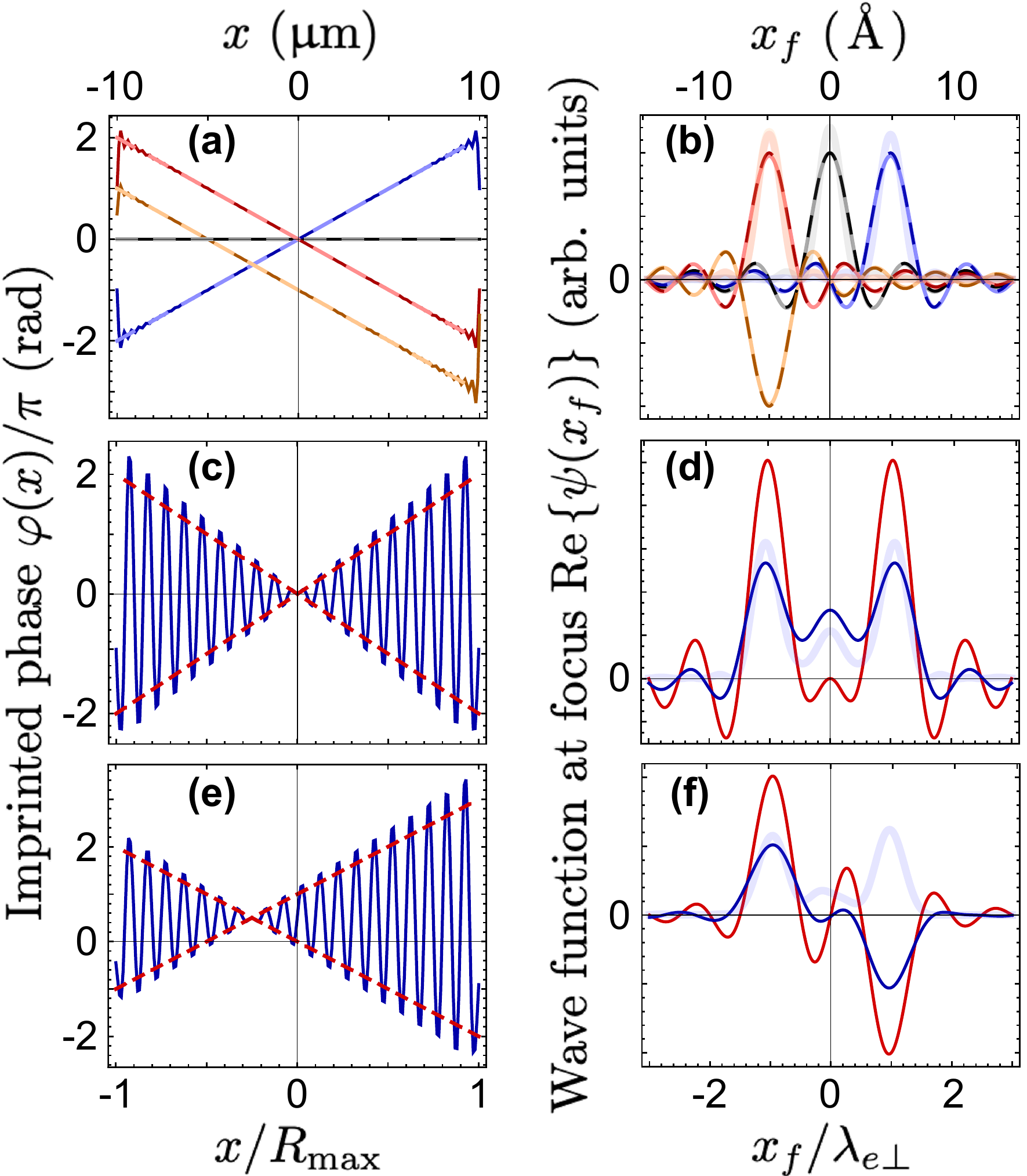}}
\caption{{\bf 1D electron focus shaping.} We plot the OFEM-imprinted electron phase (a,c,e) and the corresponding wave function at the focal plane (b,d,f). Dashed curves in (a,b) and red curves in (c-f) correspond to ideal target profiles, while solid curves in (a,b) and blue curves in (c-f) stand for the result obtained by introducing optical diffraction in the OFEM illumination. We consider a linear phase variation (a) leading to single-peak wave functions (b), as well as more complex phase patterns (c,e) producing symmetric (d) and antisymmetric (f) double-peak wave functions. We take a ratio of the objective-lens semi-aperture to the light wavelength $R_{\rm max}/\lambda_0=12.5$. The in-plane OFEM and focal coordinates $x$ and $x_f$ are normalized to $R_{\rm max}$ and the projected electron Abbe wavelength $\lambda_{e\perp}=\lambda_e/{\rm NA}$, respectively, where $\lambda_e=2\pi/q_0$ is the electron wavelength and ${\rm NA}=R_{\rm max}/(z_f-z_L)$ is the microscope numerical aperture. The electron probability density $|\psi|^2$ is shown as color-matching thick-light curves in (b,d,f). Upper horizontal scales correspond to 60\,keV electrons, $R_{\rm max}=10\,\upmu$m, and ${\rm NA}=0.01$.}
\label{Fig2}
\end{figure}
$\varphi(\Rb)$ (rad)

We explore this strategy in Fig.\ \ref{Fig2}, where the left panels present the OFEM phase, whereas the right ones show the corresponding color-matched wave functions obtained by inserting that phase into Eq.\ (\ref{psifinal}) without aberrations ($\chi=0$) and with the integral over $\theta_y$ yielding a trivial overall constant factor. Broken curves in Fig.\ \ref{Fig2}(a,b) and red curves in Fig.\ \ref{Fig2}(c-f) stand for target profiles, whereas the rest of the curves are obtained by accounting for optical diffraction [i.e., by transforming the target phase as prescribed by Eq.\ (\ref{phifinal})]. In-plane OFEM and focal coordinates $x$ and $x_f$ are normalized as explained in the caption, thus defining universal curves for a specific choice of the ratio between the objective aperture radius and the light wavelength, $R_{\rm max}/\lambda_0=12.5$. Linear phase profiles [Fig.\ \ref{Fig2}(a)], which are well reproduced by diffraction-limited illumination, give rise to peaked electron wave functions [Fig.\ \ref{Fig2}(b)] centered at positions $x_f=(A/2\pi)\lambda_{e\perp}$ that depend on the slope of the phase $\varphi=Ax/R_{\rm max}+B$, with the offset value $B$ determining the focal peak phase. The situation is more complicated when aiming to produce two electron peaks, which can be achieved with an intermitent phase profile that combines two different slopes, either without [Fig.\ \ref{Fig2}(c,d)] or with [Fig.\ \ref{Fig2}(e,f)] offset to generate symmetric or antisymmetric wave functions, respectively. Light diffraction reduces the contrast of the obtained focal shapes, but still tolerates well-defined intensity peaks [Fig.\ \ref{Fig2}(b,d,f), light curves], which become sharper when $R_{\rm max}/\lambda_0$ is increased (see Fig.\ \ref{FigS1}).

\begin{figure}
\centering{\includegraphics[width=0.45\textwidth]{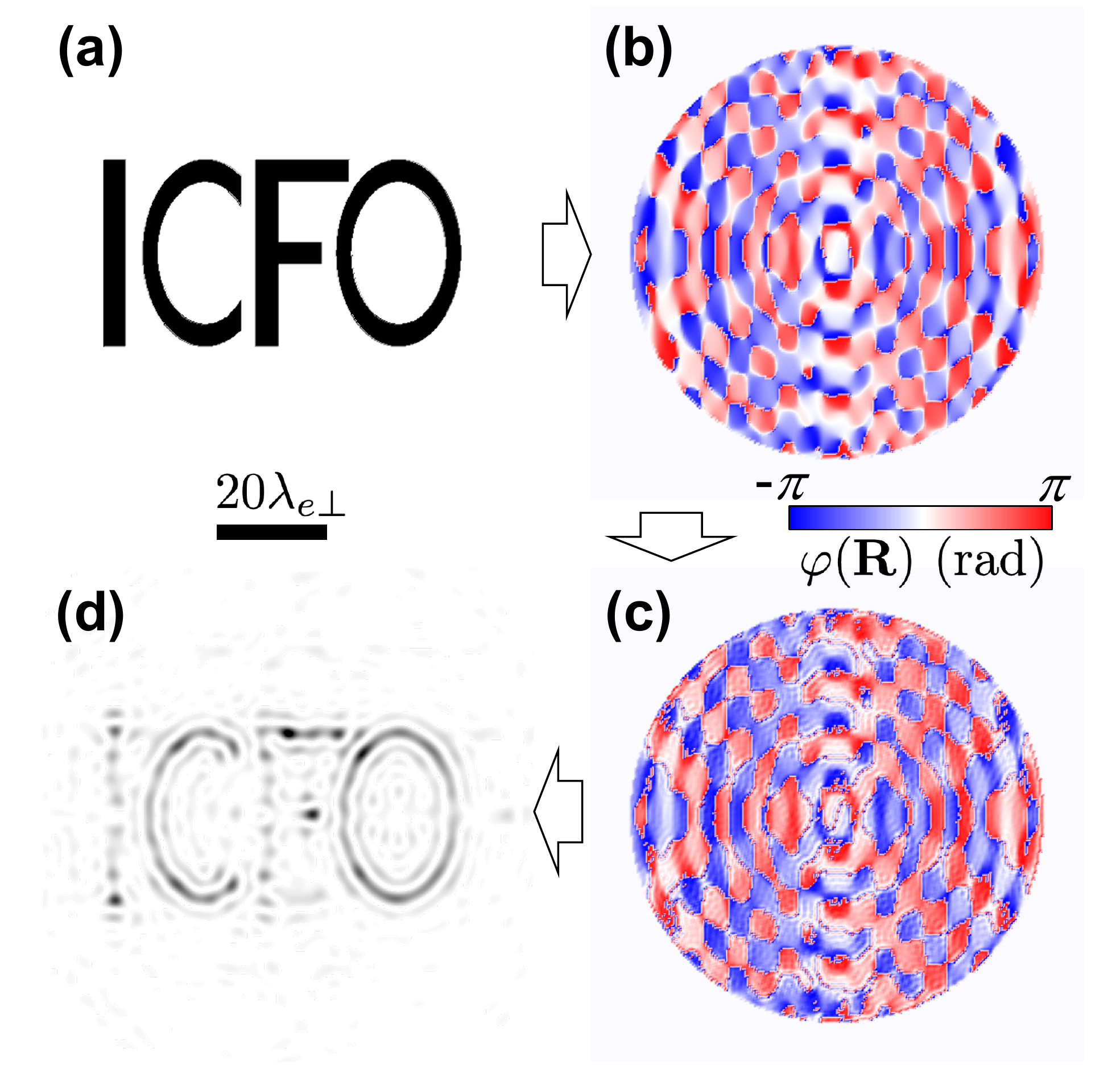}}
\caption{{\bf 2D electron focus shaping.} (a) Designated focal shape. (b) Phase of the the Fourier transform of (a) (target phase). (c) Diffraction-limited phase to be imprinted by OFEM assuming $R_{\rm max}/\lambda_0=30$. (d) Finally obtained focal profile. The bar shows the scale of (a,b) in units of $\lambda_{e\perp}$. The phase plots have a radius $R_{\rm max}$.}
\label{Fig3}
\end{figure}


For actual two-dimensional beams, using consolidated results of image compression theory \cite{HLO1980,AT10}, we can find approximate contour spot profiles by setting the OFEM phase to the argument of the Fourier transform of solid shapes filling those contours. This is illustrated in Fig.\ \ref{Fig3}, where panel (b) represents the phase of the object in (a), while panel (c) is the actual diffraction-limited phase obtained by convoluting (b) with a point spread function $\propto J_1(2\pi R/\lambda_0)/R$ (see Appendix\ \ref{secS5}), which produces a blurred, but still discernible electron focal image.

\section{Conclusion}

In brief, shaped optical fields can modulate the electron wave function in free space to produce on-demand e-beam focal profiles. The required light intensities are reachable using pulsed illumination, currently available in ultrafast electron microscope setups. We have illustrated this idea with simple examples of target and optical profiles, but a higher degree of control over the transverse electron wave function should benefit from machine learning techniques \cite{SOJ20} for the well-defined problem of finding the optimum light beam angular profile that better fits the desired e-beam spot shape. In combination with spatiotemporal light shaping, the proposed OFEM element should enable the exploration of nanoscale nonlocal correlations in sample dynamics.

\section*{Acknowledgments}

We thank Mathieu Kociak and Ido Kaminer for helpful and enjoyable discussions. This work has been supported in part by ERC (Advanced Grant 789104-eNANO), the Spanish MINECO (MAT2017-88492-R and SEV2015-0522), the Catalan CERCA Program, and Fundaci\'{o} Privada Cellex.

\begin{widetext}

\appendix

\section{Derivation of equation\ (\ref{psifinal})}
\label{secS1}

We start from Eq.\ (\ref{psi1}),
\begin{align}
\psi_{z_f}(\Rb_f)\propto\int d^2\Rb\;\ee^{\ii q_0|\Rb_f-\Rb|^2/2(z_f-z_L)}\mathcal{F}(\Rb)\psi_{z_L}(\Rb),
\label{eq2}
\end{align}
where we have inserted a transmission factor $\mathcal{F}(\Rb')=\Theta(R_{\rm max}-R)\ee^{-\ii q_0 R^2/2f}$ associated with a focusing lens of focal distance $f$ and aperture radius $R_{\rm max}$ placed at the $z=z_L$ plane. Specifying the incident electron as a spherical wave $\psi_{z_L}(\Rb)\propto\ee^{\ii q_0R^2/2(z_L-z_{\rm xo})}$ emanating from the crossover point at $z=z_{\rm xo}$ (see Fig.\ \ref{Fig1}), Eq.\ (\ref{eq2}) reduces to
\begin{align}
\psi_{z_f}(\Rb_f)\propto\int_{R<R_{\rm max}} d^2\Rb\;
\ee^{-\ii q_0\Rb\cdot\Rb'/(z_f-z_L)}
\ee^{\ii q_0R^2\Delta/2},
\label{eq22}
\end{align}
where $\Delta=1/(z_f-z_L)+1/(z_L-z_{\rm xo})-1/f$. In Eq.\ (\ref{eq22}), we dismiss an overall factor $\ee^{\ii q_0R_f^2/2(z_f-z_L)}$, which introduces a negligible phase modulation across the focal spot [e.g., for 60\,keV electrons ($q_0\approx1291/$nm) and $z_f-z_L\sim f=1\,$mm, a change in $R_f$ by 1\,nm produces a phase shift $q_0R_f^2/2(z_f-z_L)\sim0.6\,$mrad]. This expression directly becomes Eq.\ (\ref{psi1}) by defining the 2D angular coordinate $\vec{\theta}=\Rb/(z_f-z_L)$ and inserting in the integrand two phase correction factors due to aberrations and the OFEM, represented through $\ee^{\ii\chi(\vec{\theta})}$ and $\ee^{\ii\varphi(\Rb)}$, respectively.

\section{Phase imprinted by paraxial optical vortex beams}
\label{secS2}

The electric field associated with an external light beam propagating along positive $z$ directions admits the general expression
\begin{align}
\Eb(\rb)=\sum_{\sigma={\rm s,p}}\int\frac{d^2\kperpb}{(2\pi)^2}
\,\beta_{\kperpb\sigma}\,\eh_{\kperpb\sigma}\,\ee^{\ii\kperpb\cdot\Rb+\ii k_zz},
\label{Egeneral}
\end{align}
where we sum over polarizations $\sigma\!=$s,\,p and integrate over wave vectors $\kperpb=(k_x,k_y)$ within the light cone $\kperp<k_0=\omega/c$, $k_z=\sqrt{k_0^2-\kperp^2}$ is the light wave vector component along $z$, $\eh_{\kperpb{\rm s}}=(1/\kperp)(-k_y\xx+k_x\yy)$ and $\eh_{\kperpb{\rm p}}=(1/k_0\kperp)(k_z\kperpb-\kperp^2\zz)$ are unit polarization vectors, and $\beta_{\kperpb\sigma}$ are  expansion coefficients. We now consider beams with a well-defined azimuthal angular momentum number $m$ by setting $\beta_{\kperpb\sigma}=\beta_{\kperp\sigma}\ee^{\ii m\varphi_{\kperpb}}$. Inserting these coefficients in Eq.\ (\ref{Egeneral}) and carrying out the $\varphi_{\kperpb}$ integral, we find
\begin{align}
\Eb(\rb)=\frac{\ii^m}{2\pi}\int_0^{k_0}\kperp\,d\kperp\,\left[
\ii \beta_{\kperpb{\rm s}}\,\Eb_{k_zm{\rm s}}(\rb)
-\beta_{\kperpb{\rm p}}\,\Eb_{k_zm{\rm p}}(\rb)
\right],
\label{Ebessel}
\end{align}
where
\begin{subequations}
\begin{align}
\Eb_{k_zm{\rm s}}(\rb)&=\left[\frac{\ii m}{\kperp R} J_m(\kperp R)\,\RR
- J_m'(\kperp R)\,\hat{\bf \varphi}\right] \ee^{\ii m \varphi} \ee^{\ii k_z z}, \nonumber\\
\Eb_{k_zm{\rm p}}(\rb)&=\frac{k_z}{k_0}\left[\ii J_m'(\kperp R)\,\RR
- \frac{m}{\kperp R} J_m(\kperp R)\,\hat{\bf \varphi}
+\frac{\kperp}{k_z} J_m(\kperp R)\,\zz\right] \ee^{\ii m \varphi} \ee^{\ii k_z z} \nonumber
\end{align}
\end{subequations}
are Bessel beams. We now evaluate Eq.\ (\ref{phase}) using the field of Eq.\ (\ref{Ebessel}), which leads to
\begin{align}
\varphi(\Rb)&=\frac{-\alpha c}{4\pi\me v\gamma\omega^2}
\int_0^{k_0}\kperp k_zd\kperp \label{phibessel}\\
&\times\left[
\left|\beta_{\kperp{\rm s}}+(\ii k_z/k_0)\beta_{\kperp{\rm p}}\right|^2J_{m-1}^2(\kperp R)
+\left|\beta_{\kperp{\rm s}}-(\ii k_z/k_0)\beta_{\kperp{\rm p}}\right|^2J_{m+1}^2(\kperp R)
+(\kperp^2/\gamma^2k_0^2)|\beta_{\kperp{\rm p}}|^2J_m^2(\kperp R)
\right].
\nonumber
\end{align}
For optical paraxial beams constructed from a limited range of transverse wave vectors $\kperp\lesssim k_0\theta_L$, where $\theta_L$ is the divergence half-angle, with $\beta_{\kperp\sigma}\equiv\beta_\sigma$ taken to be constant within that range, we use the approximation $J_m(\theta)\approx\theta^m/2^mm!$ for $|\theta|\ll1$ to write the lowest-order contribution to Eq.\ (\ref{phibessel}) as
\begin{align}
\varphi(\Rb)\approx\frac{-m\alpha(k_0\theta_L)^{2m}}{2\pi4^mm!^2\me v\gamma\omega}\,
\left|\beta_{\rm s}+\ii\beta_{\rm p}\right|^2\,R^{2(m-1)}.
\nonumber
\end{align}
This approximation is valid for small arguments of the Bessel functions, that is, $R\ll1/k_0\theta_0=\lambda_0/2\pi\theta_L$. For a light wavelength $\lambda_0=500\,$nm and a typical objective lens radius $R_{\rm max}\sim30\,\upmu$m, this imposes an upper limit on the divergence angle of the optical beam $\theta_L\ll0.15^\circ$.

The power carried by the light beam can be obtained by integrating the Poynting vector associated with the field of Eq.\ (\ref{Egeneral}). We find\begin{align}
\mathcal{P}=\frac{c^2}{8\pi^3\omega}\sum_{\sigma={\rm s,p}}\int d^2\kperpb
\,k_z\,\left|\beta_{\kperpb\sigma}\right|^2\approx\frac{\theta_L^2\omega^2}{8\pi^2c}\left(|\beta_{\rm s}|^2+|\beta_{\rm p}|^2\right),
\label{P}
\end{align}
where the rightmost approximation corresponds to the paraxial beam considered above. When the external light is made of only p or s polarization components, we can use Eq.\ (\ref{P}) to recast the phase as
\begin{align}
\varphi(\Rb)\approx\frac{-\pi\alpha m\,\mathcal{P}}{m!^2\me cv\gamma\omega}\,\left(k_0\theta_LR/2\right)^{2(m-1)},
\label{phaseR}
\end{align}
which is obviously proportional to the beam power $\mathcal{P}$. Interestingly, by setting $m=3$ we have $\varphi(\Rb)\propto R^4$, which has the same radial scaling as the phase associated with spherical aberration in the electron beam.

\section{Effective length of interaction for an $m=1$ paraxial light beam}
\label{secS3}

For $m=1$, Eq.\ (\ref{phaseR}) is independent of $R$ and $\theta_L$ within the paraxial approximation. This allows us to estimate the power needed to obtain a phase shift of $2\pi$ in the electron as $\mathcal{P}=2\me cv\gamma\omega/\alpha$. Now, from Eq.\ (\ref{phase}), we can roughly estimate the phase shift as
\begin{align}
\varphi\sim-2\pi\alpha I_0L/\me v\gamma\omega^2,
\label{varphiL}
\end{align}
where the focal field intensity is absorbed in the light intensity $I_0=(c/2\pi)|\Eb(0)|^2$ and $L$ is an effective light-electron interaction length that depends on the light focusing conditions. In particular, we can find $L$ for the paraxial beams considered above by first calculating the field intensity from Eq.\ (\ref{Ebessel}) for $\rb=0$ and $m=1$; we find $|\Eb(0)|^2\approx(k_0^4\theta_L^4/32\pi^2)\left|\beta_{\rm s}+\ii\beta_{\rm p}\right|^2$, which permits writing the beam power in Eq.\ (\ref{P}) (for either s or p polarization) as $\mathcal{P}=8\pi c I_0/\omega^3\theta_L^2$, and from here and Eq.\ (\ref{phaseR}), we have $\varphi=-8\pi^2\alpha c I_0/\me v\gamma\omega^3\theta_L^2$. Comparing this expression with Eq.\ (\ref{varphiL}), we obtain $L=2\lambda_0/\theta_L^2$.

\section{Derivation of equation\ (\ref{phifinal})}
\label{secS4}

The electric field $\Eb(\rb)=E(x,z)\yy$ of a light beam propagating along positive $z$ directions and having translational invariance along $y$ can be regarded as a combination of plane waves of wave vectors $(k_x,k_z=\sqrt{k_0-k_x^2})$. More precisely, we can write
\begin{align}
E(x,z)=\int_{-k_0}^{k_0}\frac{dk_x}{2\pi}\,\ee^{\ii(k_xx+k_zz)}\,\beta_{k_x}
\label{Ex}
\end{align}
in terms of expansion coefficients $\beta_{k_x}$. Inserting this expression into Eq.\ (\ref{phase}) and performing the $z$ integral, we find
\begin{align}
\varphi(x)=\frac{-1}{2\pi\mathcal{M}\omega^2}\int_{-k_0}^{k_0}dk_x\int_{-k_0}^{k_0}dk'_x\,\ee^{\ii(k_x-k'_x)x}\,\beta_{k_x}\beta^*_{k'_x}\,\delta(k_z-k'_z).
\label{phixbeta}
\end{align}
Now, the $\delta$ function, which contributes only at $k_x=\pm k'_x$, can be recast as
\begin{align}
\delta(k_z-k'_z)=\frac{k_z}{|k_x|}\left[\delta(k_x-k'_x)+\delta(k_x+k'_x)\right],
\label{delta}
\end{align}
thus permitting us to carry out the $k'_x$ integral to find
\begin{align}
\varphi(x)=\varphi_0-\frac{1}{2\pi\mathcal{M}\omega^2}\int_{-k_0}^{k_0}dk_x\,\frac{k_z}{|k_x|}\,\ee^{2\ii k_xx}\,\beta_{k_x}\beta^*_{-k_x},
\label{phix}
\end{align}
where $\varphi_0=(-1/2\pi\mathcal{M}\omega^2)\int_{-k_0}^{k_0}dk_x\,(k_z/|k_x|)\,|\beta_{k_x}|^2$ is an overall phase that we dismiss because it is independent of $x$ and does not affect the electron intensity profile. Given a target OFEM phase $\varphi^{\rm target}(x)$, although the range of $k_x$ integration is not infinite, we can approximate the light beam coefficients as the inverse Fourier transform
\begin{align}
\beta_{k_x}\beta_{-k_x}^*\approx-2\mathcal{M}\omega^2\frac{|k_x|}{k_z}\int dx\,\ee^{-2\ii k_xx}\,\varphi^{\rm target}(x).
\label{betabeta}
\end{align}
A specific solution of this equation can be found by setting $\beta_{k_x}=\beta^*_{-k_x}$, which renders $\beta_{k_x}$ as the square root of the right-hand side of Eq.\ (\ref{betabeta}). For any solution, by inserting Eq.\ (\ref{betabeta}) into Eq.\ (\ref{phix}), we find
\begin{align}
\varphi(x)=\frac{1}{\pi}\int_{-R_{\rm max}}^{R_{\rm max}}\!\!dx'\,\frac{\sin\left[2k_0(x-x')\right]}{x-x'}\varphi^{\rm target}(x'),
\nonumber
\end{align}
which is Eq.\ (\ref{phifinal}). The actual phase $\varphi(x)$ that can be imprinted with a light beam in the OFEM element [i.e., taking into account the finite range of wave vectors $|k_x|\le k_0$ contributing to Eq.\ (\ref{Ex})] is the target phase convoluted with the 1D point spread function $\sin(2k_0x)/x$.

\section{Synthesis of 2D phase profiles}
\label{secS5}

In the analysis of Fig.\ \ref{Fig3}, we start with a black-and-white target intensity profile [e.g., $I(\Rb_f)=1$ and 0 in the black and white areas of Fig.\ \ref{Fig3}(a)] and perform the fast Fourier transform (FFT) \cite{PTV92} to compute $I_\Qb=\int(d^2\Rb_f/4\pi^2)\,I(\Rb_f)\,\ee^{\ii\Qb\cdot\Rb_f}$, where $\Qb=q_0\Rb/(z_f-z_L)$ [see definitions of different variables in Eq.\ (\ref{psifinal})]. We then retain only the phase of $I_\Qb$, which is known to capture the contour of the designated intensity profile \cite{HLO1980,AT10}, so we regard it as the target phase $\varphi^{\rm target}(\Rb)={\rm arg}\{I_\Qb\}$ to be delivered by the OFEM [Fig.\ \ref{Fig3}(b)]. By analogy to the 1D profile analysis in Sec.\ \ref{secS4}, we introduce the effects of light diffraction in a phenomenological way by first Fourier transforming $\varphi_{\kparb}=\int_{R<R_{\rm max}}(d^2\Rb/4\pi^2)\,\varphi^{\rm target}(\Rb)\,\ee^{-\ii\kparb\cdot\Rb}$, where $\kparb=(k_x,k_y)$ is the light wave vector component parallel to the OFEM plane and the integral extends over the objective lens aperture defined by $R<R_{\rm max}$; in a second step, we transform back to $\varphi(\Rb)=\int_{\kpar<k_0} d^2\kparb\,\varphi_{\kb}\,\ee^{\ii\kparb\cdot\Rb}$, only retaining diffraction-limited components $\kpar<k_0$ [Fig.\ \ref{Fig3}(c)]; this procedure is equivalent to calculating $\varphi(\Rb)$ as the convolution of $\varphi^{\rm target}(\Rb)$ with the 2D point spread function $k_0 J_1(k_0R)/2\pi R$. We finally obtain the actual spot profile [Fig.\ \ref{Fig3}(d)] by evaluating Eq.\ (\ref{psifinal}) without aberrations ($\chi=0$) at the focal plane ($\Delta=0$).

Incidentally, apart from an overall normalization factor, we find that the light wavelength $\lambda_0$, the electron wave vector $q_0$ (or equivalently, the electron energy $E_0$), and the in-plane focal and OFEM coordinates $\Rb_f$ and $\Rb$ enter the analysis presented in this work only through the combinations $q_0\Rb\cdot\Rb_f/(z_f-z_L)=2\pi\,(\Rb/R_{\rm max})\,(\Rb_f/\lambda_{e\perp})$ [e.g., see Eq.\ (\ref{psifinal})] and $\Rb\cdot\kparb=\Rb/R_{\rm max}=2\pi\,(\kparb/k_0)\,(\Rb/R_{\rm max})\,(R_{\rm max}/\lambda_0)$, where $\kparb/k_0$ is the in-plane projection of the unit vector indicating the incident light direction, whereas $\lambda_{e\perp}=\lambda_e/{\rm NA}$ is the projected focal-plane electron wavelength defined as the ratio between the de Broglie wavelength $\lambda_e=2\pi/q_0$ and the numerical aperture of the objective lens ${\rm NA}=R_{\rm max}/(z_f-z_L)$. By normalizing $\Rb_f$ and $\Rb$ to $\lambda_{e\perp}$ and $R_{\rm max}$, respectively, we obtain universal curves that only depend on the ratio $R_{\rm max}/\lambda_0$ between the aperture radius of the objective lens and the light wavelength. We use this normalization in Figs.\ \ref{Fig2} and \ref{Fig3}, as well as in the additional Fig.\ \ref{FigS1}.

\section{Evaluation of the OFEM phase from the incident light field amplitude}
\label{secS6}

For completeness, we present a generalization of Eqs.\ (\ref{Ex}) and (\ref{phixbeta}) to 2D beams, for which the incident field can be expressed in general as indicated in Eq.\ (\ref{Egeneral}). Inserting the latter into Eq.\ (\ref{phase}), performing the integral over $z$, and using a relation similar to Eq.\ (\ref{delta}), the phase reduces to
\begin{align}
\varphi(\Rb)=\frac{-1}{(2\pi)^3\mathcal{M}\omega^2}\sum_{\sigma\sigma'}\int d^2\kperpb\int d^2\kperpb'\,\ee^{\ii(\kperpb-\kperpb')\cdot\Rb}\,\beta_{\kperpb\sigma}\beta^*_{\kperpb'\sigma'}\;S_{\sigma,\sigma'}(\kperp,\phi,\phi')\;\frac{k_z}{\kperp}\,\delta(\kperp-\kperp')\,\Theta(k_0-\kperp),
\label{newfi}
\end{align}
where $\phi$ and $\phi'$ are the azimuthal angles of $\kperpb$ and $\kperpb'$, respectively, and we define
\begin{align}
S_{\sigma,\sigma'}(\kperp,\phi,\phi')=\eh_{\kperpb\sigma}\cdot\eh_{\kperpb'\sigma'}=\left\{
\begin{matrix}
\cos(\phi-\phi')& \quad\quad\quad\text{for }\sigma={\rm s},\;\sigma'={\rm s},\\
-\dfrac{k_z}{k_0}\sin(\phi-\phi')& \quad\quad\quad\text{for }\sigma={\rm s},\;\sigma'={\rm p},\\
\dfrac{k_z}{k_0}\sin(\phi-\phi')& \quad\quad\quad\text{for }\sigma={\rm p},\;\sigma'={\rm s},\\
\dfrac{k_z^2}{k_0^2}\cos(\phi-\phi')+\dfrac{\kperp^2}{k_0^2}& \quad\quad\quad\text{for }\sigma={\rm p},\;\sigma'={\rm p},
\end{matrix}
\right.
\nonumber
\end{align}
and $k_z=\sqrt{k_0^2-\kperp^2}$. Equation\ (\ref{newfi}) involves a 3D integral that needs to be evaluated over the coordinates $\Rb$ of the 2D lens aperture, thus demanding an unaffordable numerical effort consisting of $\sim N^5$ operations for $N\sim10^2-10^3$ points per dimension. A more practical way of evaluating this expression can be found by first computing its Fourier transform
\begin{align}
\varphi_{\kperpb}&=\int d^2\Rb\,\varphi(\Rb)\,\ee^{-\ii\kperpb\cdot\Rb} \nonumber\\
&=\frac{-1}{2\pi\mathcal{M}\omega^2}\sum_{\sigma\sigma'}\int d^2\kperpb'\,\beta_{\kperpb'\sigma}\beta^*_{\kperpb'-\kperpb,\sigma'}\;S_{\sigma,\sigma'}(\kperp',\phi',\phi'')\;\frac{k'_z}{\kperp'}\,\delta\left(\kperp'-|\kperpb'-\kperpb|\right)\,\Theta(k_0-\kperp'),
\nonumber
\end{align}
where $\phi''$ is the azimuthal angle of $\kperpb'-\kperpb$. Here, the $\delta$ function imposes $\kperp'=\kperp/\cos(\phi-\phi')$ and introduces a denominator given by the derivative of its argument. After some simple algebra, we find
\begin{align}
\varphi_{\kperpb}=\frac{-1}{4\pi\mathcal{M}\omega^2}\sum_{\sigma\sigma'}\int_{-\phi_0}^{\phi_0} \frac{d\phi'}{\cos^2(\phi-\phi')}\;k'_z\;\beta_{\kperpb'\sigma}\beta^*_{\kperpb'-\kperpb,\sigma'}\;S_{\sigma,\sigma'}(\kperp',\phi',\phi''),
\nonumber
\end{align}
where $\phi_0=\cos^{-1}\left(\kperp/2k_0\right)$. Together with the inverse Fourier transform $\varphi(\Rb)=(2\pi)^{-2}\int d^2\kperpb\,\varphi_{\kperpb}\,\ee^{\ii\kperpb\cdot\Rb}$, the evaluation now takes an affordable number of operations $\sim N^3\log^2N$ (i.e., a factor of $N$ for the 1D integral over $\phi'$ and the remaining factors needed for the 2D FFTs), which can be beneficial for carrying out extensive phase calculations, as needed to train machine learning algorithms for fast determination of the light coefficients $\beta_{\kperpb\sigma}$ in Eq.\ (\ref{Egeneral}).

\begin{figure}
\centering{\includegraphics[width=0.45\textwidth]{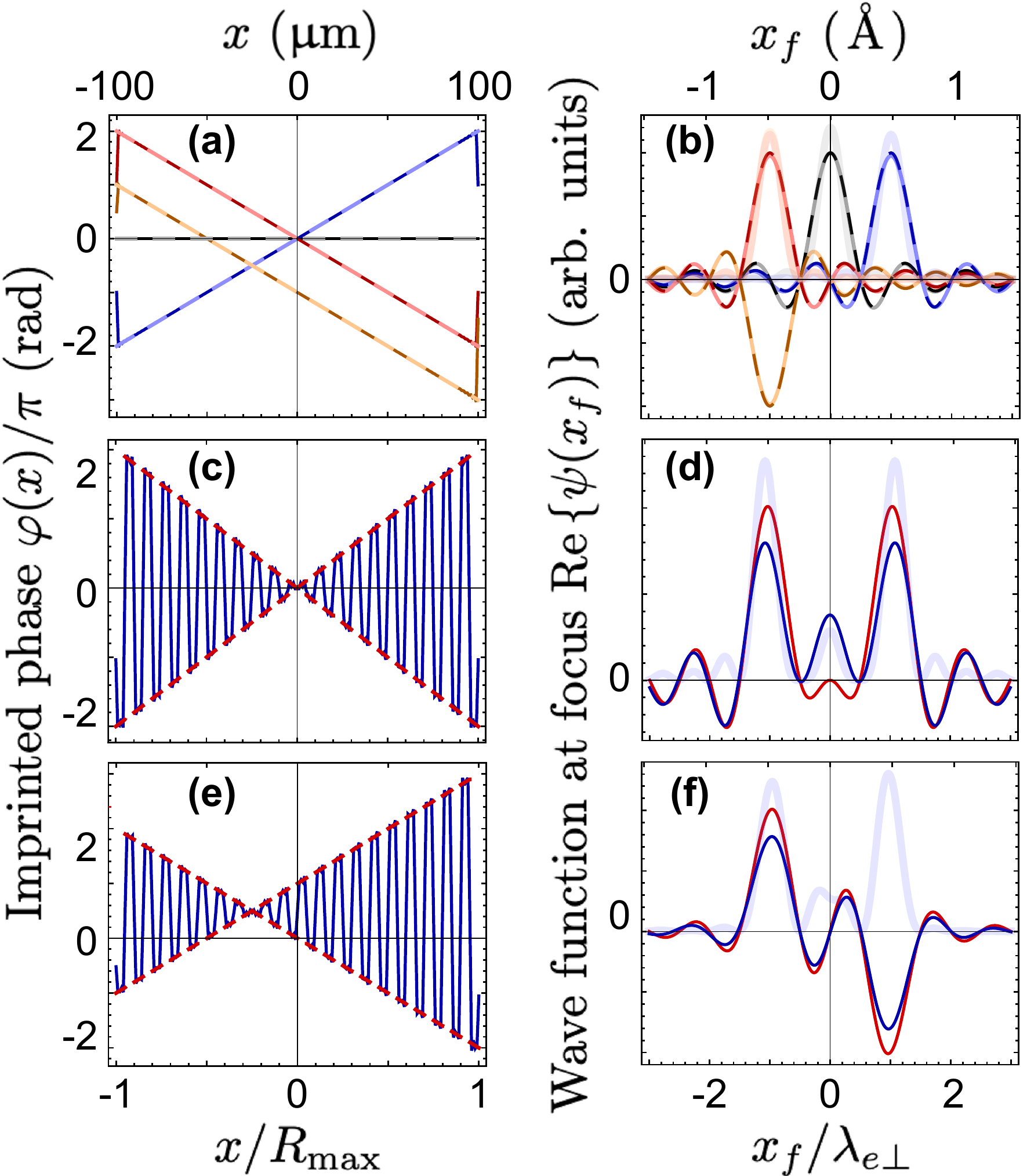}}
\caption{Same as Fig.\ \ref{Fig2}, but with a different value of the parameter $R_{\rm max}/\lambda_0=125$. Upper horizontal scales correspond to 60\,keV electrons, $R_{\rm max}=100\,\upmu$m, and ${\rm NA}=0.1$. The in-plane OFEM and focal coordinates $x$ and $x_f$ are normalized to $R_{\rm max}$ and $\lambda_{e\perp}=\lambda_e/{\rm NA}$.}
\label{FigS1}
\end{figure}

\end{widetext}


\begin{thebibliography}{67}
\expandafter\ifx\csname natexlab\endcsname\relax\def\natexlab#1{#1}\fi
\expandafter\ifx\csname bibnamefont\endcsname\relax
  \def\bibnamefont#1{#1}\fi
\expandafter\ifx\csname bibfnamefont\endcsname\relax
  \def\bibfnamefont#1{#1}\fi
\expandafter\ifx\csname citenamefont\endcsname\relax
  \def\citenamefont#1{#1}\fi
\expandafter\ifx\csname url\endcsname\relax
  \def\url#1{\texttt{#1}}\fi
\expandafter\ifx\csname urlprefix\endcsname\relax\def\urlprefix{URL }\fi
\providecommand{\bibinfo}[2]{#2}
\providecommand{\eprint}[2][]{\url{#2}}

\bibitem[{\citenamefont{Haider et~al.}(1998)\citenamefont{Haider, Rose,
  Uhlemann, Schwan, Kabius, and Urban}}]{HRU98}
\bibinfo{author}{\bibfnamefont{M.}~\bibnamefont{Haider}},
  \bibinfo{author}{\bibfnamefont{H.}~\bibnamefont{Rose}},
  \bibinfo{author}{\bibfnamefont{S.}~\bibnamefont{Uhlemann}},
  \bibinfo{author}{\bibfnamefont{E.}~\bibnamefont{Schwan}},
  \bibinfo{author}{\bibfnamefont{B.}~\bibnamefont{Kabius}}, \bibnamefont{and}
  \bibinfo{author}{\bibfnamefont{K.}~\bibnamefont{Urban}},
  \bibinfo{journal}{Ultramicroscopy} \textbf{\bibinfo{volume}{75}},
  \bibinfo{pages}{53} (\bibinfo{year}{1998}).

\bibitem[{\citenamefont{Batson et~al.}(2002)\citenamefont{Batson, Dellby, and
  Krivanek}}]{BDK02}
\bibinfo{author}{\bibfnamefont{P.~E.} \bibnamefont{Batson}},
  \bibinfo{author}{\bibfnamefont{N.}~\bibnamefont{Dellby}}, \bibnamefont{and}
  \bibinfo{author}{\bibfnamefont{O.~L.} \bibnamefont{Krivanek}},
  \bibinfo{journal}{Nature} \textbf{\bibinfo{volume}{418}},
  \bibinfo{pages}{617} (\bibinfo{year}{2002}).

\bibitem[{\citenamefont{Hawkes and Spence}(2019)}]{HS19}
\bibinfo{author}{\bibfnamefont{P.}~\bibnamefont{Hawkes}} \bibnamefont{and}
  \bibinfo{author}{\bibfnamefont{J.}~\bibnamefont{Spence}},
  \emph{\bibinfo{title}{Springer Handbook of Microscopy}}
  (\bibinfo{publisher}{Springer Nature Switzerland AG}, \bibinfo{year}{2019}).

\bibitem[{\citenamefont{Nellist et~al.}(2004)\citenamefont{Nellist, Chisholm,
  Dellby, Krivanek, Murfitt, Szilagyi, Lupini, Borisevich, {Sides Jr.}, and
  Pennycook}}]{NCD04}
\bibinfo{author}{\bibfnamefont{P.~D.} \bibnamefont{Nellist}},
  \bibinfo{author}{\bibfnamefont{M.~F.} \bibnamefont{Chisholm}},
  \bibinfo{author}{\bibfnamefont{N.}~\bibnamefont{Dellby}},
  \bibinfo{author}{\bibfnamefont{O.~L.} \bibnamefont{Krivanek}},
  \bibinfo{author}{\bibfnamefont{M.~F.} \bibnamefont{Murfitt}},
  \bibinfo{author}{\bibfnamefont{Z.~S.} \bibnamefont{Szilagyi}},
  \bibinfo{author}{\bibfnamefont{A.~R.} \bibnamefont{Lupini}},
  \bibinfo{author}{\bibfnamefont{A.}~\bibnamefont{Borisevich}},
  \bibinfo{author}{\bibfnamefont{W.~H.} \bibnamefont{{Sides Jr.}}},
  \bibnamefont{and} \bibinfo{author}{\bibfnamefont{S.~J.}
  \bibnamefont{Pennycook}}, \bibinfo{journal}{Science}
  \textbf{\bibinfo{volume}{305}}, \bibinfo{pages}{1741} (\bibinfo{year}{2004}).

\bibitem[{\citenamefont{Muller et~al.}(2008)\citenamefont{Muller, {Fitting
  Kourkoutis}, Murfitt, Song, Hwang, Silcox, Dellby, and Krivanek}}]{MKM08}
\bibinfo{author}{\bibfnamefont{D.~A.} \bibnamefont{Muller}},
  \bibinfo{author}{\bibfnamefont{L.}~\bibnamefont{{Fitting Kourkoutis}}},
  \bibinfo{author}{\bibfnamefont{M.}~\bibnamefont{Murfitt}},
  \bibinfo{author}{\bibfnamefont{J.~H.} \bibnamefont{Song}},
  \bibinfo{author}{\bibfnamefont{H.~Y.} \bibnamefont{Hwang}},
  \bibinfo{author}{\bibfnamefont{J.}~\bibnamefont{Silcox}},
  \bibinfo{author}{\bibfnamefont{N.}~\bibnamefont{Dellby}}, \bibnamefont{and}
  \bibinfo{author}{\bibfnamefont{O.~L.} \bibnamefont{Krivanek}},
  \bibinfo{journal}{Science} \textbf{\bibinfo{volume}{319}},
  \bibinfo{pages}{1073} (\bibinfo{year}{2008}).

\bibitem[{\citenamefont{Krivanek et~al.}(2014)\citenamefont{Krivanek, Lovejoy,
  Dellby, Aoki, Carpenter, Rez, Soignard, Zhu, Batson, Lagos et~al.}}]{KLD14}
\bibinfo{author}{\bibfnamefont{O.~L.} \bibnamefont{Krivanek}},
  \bibinfo{author}{\bibfnamefont{T.~C.} \bibnamefont{Lovejoy}},
  \bibinfo{author}{\bibfnamefont{N.}~\bibnamefont{Dellby}},
  \bibinfo{author}{\bibfnamefont{T.}~\bibnamefont{Aoki}},
  \bibinfo{author}{\bibfnamefont{R.~W.} \bibnamefont{Carpenter}},
  \bibinfo{author}{\bibfnamefont{P.}~\bibnamefont{Rez}},
  \bibinfo{author}{\bibfnamefont{E.}~\bibnamefont{Soignard}},
  \bibinfo{author}{\bibfnamefont{J.}~\bibnamefont{Zhu}},
  \bibinfo{author}{\bibfnamefont{P.~E.} \bibnamefont{Batson}},
  \bibinfo{author}{\bibfnamefont{M.~J.} \bibnamefont{Lagos}},
  \bibnamefont{et~al.}, \bibinfo{journal}{Nature}
  \textbf{\bibinfo{volume}{514}}, \bibinfo{pages}{209} (\bibinfo{year}{2014}).

\bibitem[{\citenamefont{Lagos et~al.}(2017)\citenamefont{Lagos, Tr\"ugler,
  Hohenester, and Batson}}]{LTH17}
\bibinfo{author}{\bibfnamefont{M.~J.} \bibnamefont{Lagos}},
  \bibinfo{author}{\bibfnamefont{A.}~\bibnamefont{Tr\"ugler}},
  \bibinfo{author}{\bibfnamefont{U.}~\bibnamefont{Hohenester}},
  \bibnamefont{and} \bibinfo{author}{\bibfnamefont{P.~E.}
  \bibnamefont{Batson}}, \bibinfo{journal}{Nature}
  \textbf{\bibinfo{volume}{543}}, \bibinfo{pages}{529} (\bibinfo{year}{2017}).

\bibitem[{\citenamefont{Hage et~al.}(2018)\citenamefont{Hage, Nicholls, Yates,
  McCulloch, Lovejoy, Dellby, Krivanek, Refson, and Ramasse}}]{HNY18}
\bibinfo{author}{\bibfnamefont{F.~S.} \bibnamefont{Hage}},
  \bibinfo{author}{\bibfnamefont{R.~J.} \bibnamefont{Nicholls}},
  \bibinfo{author}{\bibfnamefont{J.~R.} \bibnamefont{Yates}},
  \bibinfo{author}{\bibfnamefont{D.~G.} \bibnamefont{McCulloch}},
  \bibinfo{author}{\bibfnamefont{T.~C.} \bibnamefont{Lovejoy}},
  \bibinfo{author}{\bibfnamefont{N.}~\bibnamefont{Dellby}},
  \bibinfo{author}{\bibfnamefont{O.~L.} \bibnamefont{Krivanek}},
  \bibinfo{author}{\bibfnamefont{K.}~\bibnamefont{Refson}}, \bibnamefont{and}
  \bibinfo{author}{\bibfnamefont{Q.~M.} \bibnamefont{Ramasse}},
  \bibinfo{journal}{Sci.\ Adv.} \textbf{\bibinfo{volume}{4}},
  \bibinfo{pages}{eaar7495} (\bibinfo{year}{2018}).

\bibitem[{\citenamefont{Hage et~al.}(2019)\citenamefont{Hage, Kepaptsoglou,
  Ramasse, and Allen}}]{HKR19}
\bibinfo{author}{\bibfnamefont{F.~S.} \bibnamefont{Hage}},
  \bibinfo{author}{\bibfnamefont{D.~M.} \bibnamefont{Kepaptsoglou}},
  \bibinfo{author}{\bibfnamefont{Q.~M.} \bibnamefont{Ramasse}},
  \bibnamefont{and} \bibinfo{author}{\bibfnamefont{L.~J.} \bibnamefont{Allen}},
  \bibinfo{journal}{Phys.\ Rev.\ Lett.} \textbf{\bibinfo{volume}{122}},
  \bibinfo{pages}{016103} (\bibinfo{year}{2019}).

\bibitem[{\citenamefont{Hachtel et~al.}(2019)\citenamefont{Hachtel, Huang,
  Popovs, Jansone-Popova, Keum, Jakowski, Lovejoy, Dellby, Krivanek, and
  Idrobo}}]{HHP19}
\bibinfo{author}{\bibfnamefont{J.~A.} \bibnamefont{Hachtel}},
  \bibinfo{author}{\bibfnamefont{J.}~\bibnamefont{Huang}},
  \bibinfo{author}{\bibfnamefont{I.}~\bibnamefont{Popovs}},
  \bibinfo{author}{\bibfnamefont{S.}~\bibnamefont{Jansone-Popova}},
  \bibinfo{author}{\bibfnamefont{J.~K.} \bibnamefont{Keum}},
  \bibinfo{author}{\bibfnamefont{J.}~\bibnamefont{Jakowski}},
  \bibinfo{author}{\bibfnamefont{T.~C.} \bibnamefont{Lovejoy}},
  \bibinfo{author}{\bibfnamefont{N.}~\bibnamefont{Dellby}},
  \bibinfo{author}{\bibfnamefont{O.~L.} \bibnamefont{Krivanek}},
  \bibnamefont{and} \bibinfo{author}{\bibfnamefont{J.~C.}
  \bibnamefont{Idrobo}}, \bibinfo{journal}{Science}
  \textbf{\bibinfo{volume}{363}}, \bibinfo{pages}{525} (\bibinfo{year}{2019}).

\bibitem[{\citenamefont{M{\"o}llenstedt and D{\"u}ker}(1956)}]{MD1956}
\bibinfo{author}{\bibfnamefont{G.}~\bibnamefont{M{\"o}llenstedt}}
  \bibnamefont{and}
  \bibinfo{author}{\bibfnamefont{H.}~\bibnamefont{D{\"u}ker}},
  \bibinfo{journal}{Zeitschrift f{\"u}r Physik} \textbf{\bibinfo{volume}{145}},
  \bibinfo{pages}{377} (\bibinfo{year}{1956}).

\bibitem[{\citenamefont{Guzzinati et~al.}(2017)\citenamefont{Guzzinati, Beche,
  Lourenco-Martins, Martin, Kociak, and Verbeeck}}]{GBL17}
\bibinfo{author}{\bibfnamefont{G.}~\bibnamefont{Guzzinati}},
  \bibinfo{author}{\bibfnamefont{A.}~\bibnamefont{Beche}},
  \bibinfo{author}{\bibfnamefont{H.}~\bibnamefont{Lourenco-Martins}},
  \bibinfo{author}{\bibfnamefont{J.}~\bibnamefont{Martin}},
  \bibinfo{author}{\bibfnamefont{M.}~\bibnamefont{Kociak}}, \bibnamefont{and}
  \bibinfo{author}{\bibfnamefont{J.}~\bibnamefont{Verbeeck}},
  \bibinfo{journal}{Nat.\ Commun.} \textbf{\bibinfo{volume}{8}},
  \bibinfo{pages}{14999} (\bibinfo{year}{2017}).

\bibitem[{\citenamefont{B\'ech\'e et~al.}(2014)\citenamefont{B\'ech\'e,
  Van~Boxem, Van~Tendeloo, and Verbeeck}}]{BVV14}
\bibinfo{author}{\bibfnamefont{A.}~\bibnamefont{B\'ech\'e}},
  \bibinfo{author}{\bibfnamefont{R.}~\bibnamefont{Van~Boxem}},
  \bibinfo{author}{\bibfnamefont{G.}~\bibnamefont{Van~Tendeloo}},
  \bibnamefont{and} \bibinfo{author}{\bibfnamefont{J.}~\bibnamefont{Verbeeck}},
  \bibinfo{journal}{Nat.\ Phys.} \textbf{\bibinfo{volume}{10}},
  \bibinfo{pages}{26} (\bibinfo{year}{2014}).

\bibitem[{\citenamefont{Verbeeck et~al.}(2018)\citenamefont{Verbeeck,
  {B\'ech\'e}, {M\"uller-Caspary}, Guzzinati, Luong, and Hertog}}]{VBM18}
\bibinfo{author}{\bibfnamefont{J.}~\bibnamefont{Verbeeck}},
  \bibinfo{author}{\bibfnamefont{A.}~\bibnamefont{{B\'ech\'e}}},
  \bibinfo{author}{\bibfnamefont{K.}~\bibnamefont{{M\"uller-Caspary}}},
  \bibinfo{author}{\bibfnamefont{G.}~\bibnamefont{Guzzinati}},
  \bibinfo{author}{\bibfnamefont{M.~A.} \bibnamefont{Luong}}, \bibnamefont{and}
  \bibinfo{author}{\bibfnamefont{M.~D.} \bibnamefont{Hertog}},
  \bibinfo{journal}{Ultramicroscopy} \textbf{\bibinfo{volume}{190}},
  \bibinfo{pages}{58} (\bibinfo{year}{2018}).

\bibitem[{\citenamefont{Verbeeck et~al.}(2010)\citenamefont{Verbeeck, Tian, and
  Schattschneider}}]{VTS10}
\bibinfo{author}{\bibfnamefont{J.}~\bibnamefont{Verbeeck}},
  \bibinfo{author}{\bibfnamefont{H.}~\bibnamefont{Tian}}, \bibnamefont{and}
  \bibinfo{author}{\bibfnamefont{P.}~\bibnamefont{Schattschneider}},
  \bibinfo{journal}{Nature} \textbf{\bibinfo{volume}{467}},
  \bibinfo{pages}{301} (\bibinfo{year}{2010}).

\bibitem[{\citenamefont{McMorran et~al.}(2011)\citenamefont{McMorran, Agrawal,
  Anderson, Herzing, Lezec, McClelland, and Unguris}}]{MAA11}
\bibinfo{author}{\bibfnamefont{B.~J.} \bibnamefont{McMorran}},
  \bibinfo{author}{\bibfnamefont{A.}~\bibnamefont{Agrawal}},
  \bibinfo{author}{\bibfnamefont{I.~M.} \bibnamefont{Anderson}},
  \bibinfo{author}{\bibfnamefont{A.~A.} \bibnamefont{Herzing}},
  \bibinfo{author}{\bibfnamefont{H.~J.} \bibnamefont{Lezec}},
  \bibinfo{author}{\bibfnamefont{J.~J.} \bibnamefont{McClelland}},
  \bibnamefont{and} \bibinfo{author}{\bibfnamefont{J.}~\bibnamefont{Unguris}},
  \bibinfo{journal}{Science} \textbf{\bibinfo{volume}{331}},
  \bibinfo{pages}{192} (\bibinfo{year}{2011}).

\bibitem[{\citenamefont{Shiloh et~al.}(2014)\citenamefont{Shiloh, Lereah,
  Lilach, and Arie}}]{SLL14}
\bibinfo{author}{\bibfnamefont{R.}~\bibnamefont{Shiloh}},
  \bibinfo{author}{\bibfnamefont{Y.}~\bibnamefont{Lereah}},
  \bibinfo{author}{\bibfnamefont{Y.}~\bibnamefont{Lilach}}, \bibnamefont{and}
  \bibinfo{author}{\bibfnamefont{A.}~\bibnamefont{Arie}},
  \bibinfo{journal}{Ultramicroscopy} \textbf{\bibinfo{volume}{144}},
  \bibinfo{pages}{26} (\bibinfo{year}{2014}).

\bibitem[{\citenamefont{Grillo et~al.}(2017)\citenamefont{Grillo, Tavabi,
  Yucelen, Lu, Venturi, Larocque, Jin, Savenko, Gazzadi, Balboni
  et~al.}}]{GTY17}
\bibinfo{author}{\bibfnamefont{V.}~\bibnamefont{Grillo}},
  \bibinfo{author}{\bibfnamefont{A.~H.} \bibnamefont{Tavabi}},
  \bibinfo{author}{\bibfnamefont{E.}~\bibnamefont{Yucelen}},
  \bibinfo{author}{\bibfnamefont{P.-H.} \bibnamefont{Lu}},
  \bibinfo{author}{\bibfnamefont{F.}~\bibnamefont{Venturi}},
  \bibinfo{author}{\bibfnamefont{H.}~\bibnamefont{Larocque}},
  \bibinfo{author}{\bibfnamefont{L.}~\bibnamefont{Jin}},
  \bibinfo{author}{\bibfnamefont{A.}~\bibnamefont{Savenko}},
  \bibinfo{author}{\bibfnamefont{G.~C.} \bibnamefont{Gazzadi}},
  \bibinfo{author}{\bibfnamefont{R.}~\bibnamefont{Balboni}},
  \bibnamefont{et~al.}, \bibinfo{journal}{Opt.\ Express}
  \textbf{\bibinfo{volume}{25}}, \bibinfo{pages}{21851} (\bibinfo{year}{2017}).

\bibitem[{\citenamefont{Shiloh et~al.}(2018)\citenamefont{Shiloh, Remez, Lu,
  Jin, Lereah, Tavabi, Dunin-Borkowski, and Arie}}]{SRP18}
\bibinfo{author}{\bibfnamefont{R.}~\bibnamefont{Shiloh}},
  \bibinfo{author}{\bibfnamefont{R.}~\bibnamefont{Remez}},
  \bibinfo{author}{\bibfnamefont{P.-H.} \bibnamefont{Lu}},
  \bibinfo{author}{\bibfnamefont{L.}~\bibnamefont{Jin}},
  \bibinfo{author}{\bibfnamefont{Y.}~\bibnamefont{Lereah}},
  \bibinfo{author}{\bibfnamefont{A.~H.} \bibnamefont{Tavabi}},
  \bibinfo{author}{\bibfnamefont{R.~E.} \bibnamefont{Dunin-Borkowski}},
  \bibnamefont{and} \bibinfo{author}{\bibfnamefont{A.}~\bibnamefont{Arie}},
  \bibinfo{journal}{Ultramicroscopy} \textbf{\bibinfo{volume}{189}},
  \bibinfo{pages}{46} (\bibinfo{year}{2018}).

\bibitem[{\citenamefont{Kapitza and Dirac}(1933)}]{KD1933}
\bibinfo{author}{\bibfnamefont{P.~L.} \bibnamefont{Kapitza}} \bibnamefont{and}
  \bibinfo{author}{\bibfnamefont{P.~A.~M.} \bibnamefont{Dirac}},
  \bibinfo{journal}{Proc.\ Cambridge\ Philos.\ Soc.}
  \textbf{\bibinfo{volume}{29}}, \bibinfo{pages}{297} (\bibinfo{year}{1933}).

\bibitem[{\citenamefont{Freimund et~al.}(2001)\citenamefont{Freimund,
  Aflatooni, and Batelaan}}]{FAB01}
\bibinfo{author}{\bibfnamefont{D.~L.} \bibnamefont{Freimund}},
  \bibinfo{author}{\bibfnamefont{K.}~\bibnamefont{Aflatooni}},
  \bibnamefont{and} \bibinfo{author}{\bibfnamefont{H.}~\bibnamefont{Batelaan}},
  \bibinfo{journal}{Nature} \textbf{\bibinfo{volume}{413}},
  \bibinfo{pages}{142} (\bibinfo{year}{2001}).

\bibitem[{\citenamefont{Freimund and Batelaan}(2002)}]{FB02}
\bibinfo{author}{\bibfnamefont{D.~L.} \bibnamefont{Freimund}} \bibnamefont{and}
  \bibinfo{author}{\bibfnamefont{H.}~\bibnamefont{Batelaan}},
  \bibinfo{journal}{Phys.\ Rev.\ Lett.} \textbf{\bibinfo{volume}{89}},
  \bibinfo{pages}{283602} (\bibinfo{year}{2002}).

\bibitem[{\citenamefont{Batelaan}(2007)}]{B07}
\bibinfo{author}{\bibfnamefont{H.}~\bibnamefont{Batelaan}},
  \bibinfo{journal}{Rev.\ Mod.\ Phys.} \textbf{\bibinfo{volume}{79}},
  \bibinfo{pages}{929} (\bibinfo{year}{2007}).

\bibitem[{\citenamefont{{Garc\'{\i}a de Abajo}}(2010)}]{paper149}
\bibinfo{author}{\bibfnamefont{F.~J.} \bibnamefont{{Garc\'{\i}a de Abajo}}},
  \bibinfo{journal}{Rev.\ Mod.\ Phys.} \textbf{\bibinfo{volume}{82}},
  \bibinfo{pages}{209} (\bibinfo{year}{2010}).

\bibitem[{\citenamefont{Koz'ak et~al.}(2018)\citenamefont{Koz'ak,
  Sch\"onenberger, and Hommelhoff}}]{KSH18}
\bibinfo{author}{\bibfnamefont{M.}~\bibnamefont{Koz'ak}},
  \bibinfo{author}{\bibfnamefont{N.}~\bibnamefont{Sch\"onenberger}},
  \bibnamefont{and}
  \bibinfo{author}{\bibfnamefont{P.}~\bibnamefont{Hommelhoff}},
  \bibinfo{journal}{Phys.\ Rev.\ Lett.} \textbf{\bibinfo{volume}{120}},
  \bibinfo{pages}{103203} (\bibinfo{year}{2018}).

\bibitem[{\citenamefont{Koz'ak}(2019)}]{K19_2}
\bibinfo{author}{\bibfnamefont{M.}~\bibnamefont{Koz'ak}},
  \bibinfo{journal}{Phys.\ Rev.\ Lett.} \textbf{\bibinfo{volume}{123}},
  \bibinfo{pages}{203202} (\bibinfo{year}{2019}).

\bibitem[{\citenamefont{Black et~al.}(2019)\citenamefont{Black, Niedermayer,
  Miao, Zhao, Solgaard, Byer, and Leedle}}]{BNM19}
\bibinfo{author}{\bibfnamefont{D.~S.} \bibnamefont{Black}},
  \bibinfo{author}{\bibfnamefont{U.}~\bibnamefont{Niedermayer}},
  \bibinfo{author}{\bibfnamefont{Y.}~\bibnamefont{Miao}},
  \bibinfo{author}{\bibfnamefont{Z.}~\bibnamefont{Zhao}},
  \bibinfo{author}{\bibfnamefont{O.}~\bibnamefont{Solgaard}},
  \bibinfo{author}{\bibfnamefont{R.~L.} \bibnamefont{Byer}}, \bibnamefont{and}
  \bibinfo{author}{\bibfnamefont{K.~J.} \bibnamefont{Leedle}},
  \bibinfo{journal}{Phys.\ Rev.\ Lett.} \textbf{\bibinfo{volume}{123}},
  \bibinfo{pages}{264802} (\bibinfo{year}{2019}).

\bibitem[{\citenamefont{Sch\"nenberger
  et~al.}(2019)\citenamefont{Sch\"nenberger, Mittelbach, Yousefi, McNeur,
  Niedermayer, and Hommelhoff}}]{SMY19}
\bibinfo{author}{\bibfnamefont{N.}~\bibnamefont{Sch\"nenberger}},
  \bibinfo{author}{\bibfnamefont{A.}~\bibnamefont{Mittelbach}},
  \bibinfo{author}{\bibfnamefont{P.}~\bibnamefont{Yousefi}},
  \bibinfo{author}{\bibfnamefont{J.}~\bibnamefont{McNeur}},
  \bibinfo{author}{\bibfnamefont{U.}~\bibnamefont{Niedermayer}},
  \bibnamefont{and}
  \bibinfo{author}{\bibfnamefont{P.}~\bibnamefont{Hommelhoff}},
  \bibinfo{journal}{Phys.\ Rev.\ Lett.} \textbf{\bibinfo{volume}{123}},
  \bibinfo{pages}{264803} (\bibinfo{year}{2019}).

\bibitem[{\citenamefont{Howie}(1999)}]{H99}
\bibinfo{author}{\bibfnamefont{A.}~\bibnamefont{Howie}},
  \bibinfo{journal}{Inst.\ Phys.\ Conf.\ Ser.} \textbf{\bibinfo{volume}{161}},
  \bibinfo{pages}{311} (\bibinfo{year}{1999}).

\bibitem[{\citenamefont{{Garc\'{\i}a de Abajo} and Kociak}(2008)}]{paper114}
\bibinfo{author}{\bibfnamefont{F.~J.} \bibnamefont{{Garc\'{\i}a de Abajo}}}
  \bibnamefont{and} \bibinfo{author}{\bibfnamefont{M.}~\bibnamefont{Kociak}},
  \bibinfo{journal}{New\ J.\ Phys.} \textbf{\bibinfo{volume}{10}},
  \bibinfo{pages}{073035} (\bibinfo{year}{2008}).

\bibitem[{\citenamefont{Howie}(2009)}]{H09}
\bibinfo{author}{\bibfnamefont{A.}~\bibnamefont{Howie}},
  \bibinfo{journal}{Microsc.\ Microanal.} \textbf{\bibinfo{volume}{15}},
  \bibinfo{pages}{314} (\bibinfo{year}{2009}).

\bibitem[{\citenamefont{Pomarico et~al.}(2018)\citenamefont{Pomarico, Madan,
  Berruto, Vanacore, Wang, Kaminer, {Garc\'{\i}a de Abajo}, and
  Carbone}}]{paper306}
\bibinfo{author}{\bibfnamefont{E.}~\bibnamefont{Pomarico}},
  \bibinfo{author}{\bibfnamefont{I.}~\bibnamefont{Madan}},
  \bibinfo{author}{\bibfnamefont{G.}~\bibnamefont{Berruto}},
  \bibinfo{author}{\bibfnamefont{G.~M.} \bibnamefont{Vanacore}},
  \bibinfo{author}{\bibfnamefont{K.}~\bibnamefont{Wang}},
  \bibinfo{author}{\bibfnamefont{I.}~\bibnamefont{Kaminer}},
  \bibinfo{author}{\bibfnamefont{F.~J.} \bibnamefont{{Garc\'{\i}a de Abajo}}},
  \bibnamefont{and} \bibinfo{author}{\bibfnamefont{F.}~\bibnamefont{Carbone}},
  \bibinfo{journal}{ACS\ Photonics} \textbf{\bibinfo{volume}{5}},
  \bibinfo{pages}{759} (\bibinfo{year}{2018}).

\bibitem[{\citenamefont{Barwick et~al.}(2009)\citenamefont{Barwick, Flannigan,
  and Zewail}}]{BFZ09}
\bibinfo{author}{\bibfnamefont{B.}~\bibnamefont{Barwick}},
  \bibinfo{author}{\bibfnamefont{D.~J.} \bibnamefont{Flannigan}},
  \bibnamefont{and} \bibinfo{author}{\bibfnamefont{A.~H.}
  \bibnamefont{Zewail}}, \bibinfo{journal}{Nature}
  \textbf{\bibinfo{volume}{462}}, \bibinfo{pages}{902} (\bibinfo{year}{2009}).

\bibitem[{\citenamefont{{Garc\'{\i}a de Abajo}
  et~al.}(2010)\citenamefont{{Garc\'{\i}a de Abajo}, {Asenjo Garcia}, and
  Kociak}}]{paper151}
\bibinfo{author}{\bibfnamefont{F.~J.} \bibnamefont{{Garc\'{\i}a de Abajo}}},
  \bibinfo{author}{\bibfnamefont{A.}~\bibnamefont{{Asenjo Garcia}}},
  \bibnamefont{and} \bibinfo{author}{\bibfnamefont{M.}~\bibnamefont{Kociak}},
  \bibinfo{journal}{Nano\ Lett.} \textbf{\bibinfo{volume}{10}},
  \bibinfo{pages}{1859} (\bibinfo{year}{2010}).

\bibitem[{\citenamefont{Park et~al.}(2010)\citenamefont{Park, Lin, and
  Zewail}}]{PLZ10}
\bibinfo{author}{\bibfnamefont{S.~T.} \bibnamefont{Park}},
  \bibinfo{author}{\bibfnamefont{M.}~\bibnamefont{Lin}}, \bibnamefont{and}
  \bibinfo{author}{\bibfnamefont{A.~H.} \bibnamefont{Zewail}},
  \bibinfo{journal}{New\ J.\ Phys.} \textbf{\bibinfo{volume}{12}},
  \bibinfo{pages}{123028} (\bibinfo{year}{2010}).

\bibitem[{\citenamefont{Park and Zewail}(2012)}]{PZ12}
\bibinfo{author}{\bibfnamefont{S.~T.} \bibnamefont{Park}} \bibnamefont{and}
  \bibinfo{author}{\bibfnamefont{A.~H.} \bibnamefont{Zewail}},
  \bibinfo{journal}{J.\ Phys.\ Chem.\ A} \textbf{\bibinfo{volume}{116}},
  \bibinfo{pages}{11128} (\bibinfo{year}{2012}).

\bibitem[{\citenamefont{Kirchner et~al.}(2014)\citenamefont{Kirchner, Gliserin,
  Krausz, and Baum}}]{KGK14}
\bibinfo{author}{\bibfnamefont{F.~O.} \bibnamefont{Kirchner}},
  \bibinfo{author}{\bibfnamefont{A.}~\bibnamefont{Gliserin}},
  \bibinfo{author}{\bibfnamefont{F.}~\bibnamefont{Krausz}}, \bibnamefont{and}
  \bibinfo{author}{\bibfnamefont{P.}~\bibnamefont{Baum}},
  \bibinfo{journal}{Nat.\ Photon.} \textbf{\bibinfo{volume}{8}},
  \bibinfo{pages}{52} (\bibinfo{year}{2014}).

\bibitem[{\citenamefont{Piazza et~al.}(2015)\citenamefont{Piazza, Lummen,
  {Qui\~{n}onez}, Murooka, Reed, Barwick, and Carbone}}]{PLQ15}
\bibinfo{author}{\bibfnamefont{L.}~\bibnamefont{Piazza}},
  \bibinfo{author}{\bibfnamefont{T.~T.~A.} \bibnamefont{Lummen}},
  \bibinfo{author}{\bibfnamefont{E.}~\bibnamefont{{Qui\~{n}onez}}},
  \bibinfo{author}{\bibfnamefont{Y.}~\bibnamefont{Murooka}},
  \bibinfo{author}{\bibfnamefont{B.}~\bibnamefont{Reed}},
  \bibinfo{author}{\bibfnamefont{B.}~\bibnamefont{Barwick}}, \bibnamefont{and}
  \bibinfo{author}{\bibfnamefont{F.}~\bibnamefont{Carbone}},
  \bibinfo{journal}{Nat.\ Commun.} \textbf{\bibinfo{volume}{6}},
  \bibinfo{pages}{6407} (\bibinfo{year}{2015}).

\bibitem[{\citenamefont{Feist et~al.}(2015)\citenamefont{Feist, Echternkamp,
  Schauss, Yalunin, Sch\"afer, and Ropers}}]{FES15}
\bibinfo{author}{\bibfnamefont{A.}~\bibnamefont{Feist}},
  \bibinfo{author}{\bibfnamefont{K.~E.} \bibnamefont{Echternkamp}},
  \bibinfo{author}{\bibfnamefont{J.}~\bibnamefont{Schauss}},
  \bibinfo{author}{\bibfnamefont{S.~V.} \bibnamefont{Yalunin}},
  \bibinfo{author}{\bibfnamefont{S.}~\bibnamefont{Sch\"afer}},
  \bibnamefont{and} \bibinfo{author}{\bibfnamefont{C.}~\bibnamefont{Ropers}},
  \bibinfo{journal}{Nature} \textbf{\bibinfo{volume}{521}},
  \bibinfo{pages}{200} (\bibinfo{year}{2015}).

\bibitem[{\citenamefont{Lummen et~al.}(2016)\citenamefont{Lummen, Lamb,
  Berruto, LaGrange, Negro, {Garc\'{\i}a de Abajo}, McGrouther, Barwick, and
  Carbone}}]{paper282}
\bibinfo{author}{\bibfnamefont{T.~T.~A.} \bibnamefont{Lummen}},
  \bibinfo{author}{\bibfnamefont{R.~J.} \bibnamefont{Lamb}},
  \bibinfo{author}{\bibfnamefont{G.}~\bibnamefont{Berruto}},
  \bibinfo{author}{\bibfnamefont{T.}~\bibnamefont{LaGrange}},
  \bibinfo{author}{\bibfnamefont{L.~D.} \bibnamefont{Negro}},
  \bibinfo{author}{\bibfnamefont{F.~J.} \bibnamefont{{Garc\'{\i}a de Abajo}}},
  \bibinfo{author}{\bibfnamefont{D.}~\bibnamefont{McGrouther}},
  \bibinfo{author}{\bibfnamefont{B.}~\bibnamefont{Barwick}}, \bibnamefont{and}
  \bibinfo{author}{\bibfnamefont{F.}~\bibnamefont{Carbone}},
  \bibinfo{journal}{Nat.\ Commun.} \textbf{\bibinfo{volume}{7}},
  \bibinfo{pages}{13156} (\bibinfo{year}{2016}).

\bibitem[{\citenamefont{Echternkamp et~al.}(2016)\citenamefont{Echternkamp,
  Feist, Sch\"{a}fer, and Ropers}}]{EFS16}
\bibinfo{author}{\bibfnamefont{K.~E.} \bibnamefont{Echternkamp}},
  \bibinfo{author}{\bibfnamefont{A.}~\bibnamefont{Feist}},
  \bibinfo{author}{\bibfnamefont{S.}~\bibnamefont{Sch\"{a}fer}},
  \bibnamefont{and} \bibinfo{author}{\bibfnamefont{C.}~\bibnamefont{Ropers}},
  \bibinfo{journal}{Nat.\ Phys.} \textbf{\bibinfo{volume}{12}},
  \bibinfo{pages}{1000} (\bibinfo{year}{2016}).

\bibitem[{\citenamefont{Ryabov and Baum}(2016)}]{RB16}
\bibinfo{author}{\bibfnamefont{A.}~\bibnamefont{Ryabov}} \bibnamefont{and}
  \bibinfo{author}{\bibfnamefont{P.}~\bibnamefont{Baum}},
  \bibinfo{journal}{Science} \textbf{\bibinfo{volume}{353}},
  \bibinfo{pages}{374} (\bibinfo{year}{2016}).

\bibitem[{\citenamefont{Vanacore et~al.}(2016)\citenamefont{Vanacore,
  Fitzpatrick, and Zewail}}]{VFZ16}
\bibinfo{author}{\bibfnamefont{G.~M.} \bibnamefont{Vanacore}},
  \bibinfo{author}{\bibfnamefont{A.~W.~P.} \bibnamefont{Fitzpatrick}},
  \bibnamefont{and} \bibinfo{author}{\bibfnamefont{A.~H.}
  \bibnamefont{Zewail}}, \bibinfo{journal}{Nano\ Today}
  \textbf{\bibinfo{volume}{11}}, \bibinfo{pages}{228} (\bibinfo{year}{2016}).

\bibitem[{\citenamefont{Koz\'ak et~al.}(2017)\citenamefont{Koz\'ak, McNeur,
  Leedle, Deng, Sch\"onenberger, Ruehl, Hartl, Harris, Byer, and
  Hommelhoff}}]{KML17}
\bibinfo{author}{\bibfnamefont{M.}~\bibnamefont{Koz\'ak}},
  \bibinfo{author}{\bibfnamefont{J.}~\bibnamefont{McNeur}},
  \bibinfo{author}{\bibfnamefont{K.~J.} \bibnamefont{Leedle}},
  \bibinfo{author}{\bibfnamefont{H.}~\bibnamefont{Deng}},
  \bibinfo{author}{\bibfnamefont{N.}~\bibnamefont{Sch\"onenberger}},
  \bibinfo{author}{\bibfnamefont{A.}~\bibnamefont{Ruehl}},
  \bibinfo{author}{\bibfnamefont{I.}~\bibnamefont{Hartl}},
  \bibinfo{author}{\bibfnamefont{J.~S.} \bibnamefont{Harris}},
  \bibinfo{author}{\bibfnamefont{R.~L.} \bibnamefont{Byer}}, \bibnamefont{and}
  \bibinfo{author}{\bibfnamefont{P.}~\bibnamefont{Hommelhoff}},
  \bibinfo{journal}{Nat.\ Commun.} \textbf{\bibinfo{volume}{8}},
  \bibinfo{pages}{14342} (\bibinfo{year}{2017}).

\bibitem[{\citenamefont{Feist et~al.}(2017)\citenamefont{Feist, Bach,
  N.~Rubiano~{da Silva}, M\"{a}ller, Priebe, Domr\"{a}se, Gatzmann, Rost,
  Schauss, Strauch et~al.}}]{FBR17}
\bibinfo{author}{\bibfnamefont{A.}~\bibnamefont{Feist}},
  \bibinfo{author}{\bibfnamefont{N.}~\bibnamefont{Bach}},
  \bibinfo{author}{\bibfnamefont{T.~D.} \bibnamefont{N.~Rubiano~{da Silva}}},
  \bibinfo{author}{\bibfnamefont{M.}~\bibnamefont{M\"{a}ller}},
  \bibinfo{author}{\bibfnamefont{K.~E.} \bibnamefont{Priebe}},
  \bibinfo{author}{\bibfnamefont{T.}~\bibnamefont{Domr\"{a}se}},
  \bibinfo{author}{\bibfnamefont{J.~G.} \bibnamefont{Gatzmann}},
  \bibinfo{author}{\bibfnamefont{S.}~\bibnamefont{Rost}},
  \bibinfo{author}{\bibfnamefont{J.}~\bibnamefont{Schauss}},
  \bibinfo{author}{\bibfnamefont{S.}~\bibnamefont{Strauch}},
  \bibnamefont{et~al.}, \bibinfo{journal}{Ultramicroscopy}
  \textbf{\bibinfo{volume}{176}}, \bibinfo{pages}{63} (\bibinfo{year}{2017}).

\bibitem[{\citenamefont{Priebe et~al.}(2017)\citenamefont{Priebe, Rathje,
  Yalunin, Hohage, Feist, Sch\"{a}fer, and Ropers}}]{PRY17}
\bibinfo{author}{\bibfnamefont{K.~E.} \bibnamefont{Priebe}},
  \bibinfo{author}{\bibfnamefont{C.}~\bibnamefont{Rathje}},
  \bibinfo{author}{\bibfnamefont{S.~V.} \bibnamefont{Yalunin}},
  \bibinfo{author}{\bibfnamefont{T.}~\bibnamefont{Hohage}},
  \bibinfo{author}{\bibfnamefont{A.}~\bibnamefont{Feist}},
  \bibinfo{author}{\bibfnamefont{S.}~\bibnamefont{Sch\"{a}fer}},
  \bibnamefont{and} \bibinfo{author}{\bibfnamefont{C.}~\bibnamefont{Ropers}},
  \bibinfo{journal}{Nat.\ Photon.} \textbf{\bibinfo{volume}{11}},
  \bibinfo{pages}{793} (\bibinfo{year}{2017}).

\bibitem[{\citenamefont{Vanacore et~al.}(2018)\citenamefont{Vanacore, Madan,
  Berruto, Wang, Pomarico, Lamb, McGrouther, Kaminer, Barwick, {Garc\'{\i}a de
  Abajo} et~al.}}]{paper311}
\bibinfo{author}{\bibfnamefont{G.~M.} \bibnamefont{Vanacore}},
  \bibinfo{author}{\bibfnamefont{I.}~\bibnamefont{Madan}},
  \bibinfo{author}{\bibfnamefont{G.}~\bibnamefont{Berruto}},
  \bibinfo{author}{\bibfnamefont{K.}~\bibnamefont{Wang}},
  \bibinfo{author}{\bibfnamefont{E.}~\bibnamefont{Pomarico}},
  \bibinfo{author}{\bibfnamefont{R.~J.} \bibnamefont{Lamb}},
  \bibinfo{author}{\bibfnamefont{D.}~\bibnamefont{McGrouther}},
  \bibinfo{author}{\bibfnamefont{I.}~\bibnamefont{Kaminer}},
  \bibinfo{author}{\bibfnamefont{B.}~\bibnamefont{Barwick}},
  \bibinfo{author}{\bibfnamefont{F.~J.} \bibnamefont{{Garc\'{\i}a de Abajo}}},
  \bibnamefont{et~al.}, \bibinfo{journal}{Nat.\ Commun.}
  \textbf{\bibinfo{volume}{9}}, \bibinfo{pages}{2694} (\bibinfo{year}{2018}).

\bibitem[{\citenamefont{Morimoto and Baum}(2018{\natexlab{a}})}]{MB18}
\bibinfo{author}{\bibfnamefont{Y.}~\bibnamefont{Morimoto}} \bibnamefont{and}
  \bibinfo{author}{\bibfnamefont{P.}~\bibnamefont{Baum}},
  \bibinfo{journal}{Phys.\ Rev.\ A} \textbf{\bibinfo{volume}{97}},
  \bibinfo{pages}{033815} (\bibinfo{year}{2018}{\natexlab{a}}).

\bibitem[{\citenamefont{Morimoto and Baum}(2018{\natexlab{b}})}]{MB18_2}
\bibinfo{author}{\bibfnamefont{Y.}~\bibnamefont{Morimoto}} \bibnamefont{and}
  \bibinfo{author}{\bibfnamefont{P.}~\bibnamefont{Baum}},
  \bibinfo{journal}{Nat.\ Phys.} \textbf{\bibinfo{volume}{14}},
  \bibinfo{pages}{252} (\bibinfo{year}{2018}{\natexlab{b}}).

\bibitem[{\citenamefont{Das et~al.}(2019)\citenamefont{Das, Blazit, Tenc\'e,
  Zagonel, Auad, Lee, Ling, Losquin, C.~Colliex, {Garc\'{\i}a de Abajo}
  et~al.}}]{paper325}
\bibinfo{author}{\bibfnamefont{P.}~\bibnamefont{Das}},
  \bibinfo{author}{\bibfnamefont{J.~D.} \bibnamefont{Blazit}},
  \bibinfo{author}{\bibfnamefont{M.}~\bibnamefont{Tenc\'e}},
  \bibinfo{author}{\bibfnamefont{L.~F.} \bibnamefont{Zagonel}},
  \bibinfo{author}{\bibfnamefont{Y.}~\bibnamefont{Auad}},
  \bibinfo{author}{\bibfnamefont{Y.~H.} \bibnamefont{Lee}},
  \bibinfo{author}{\bibfnamefont{X.~Y.} \bibnamefont{Ling}},
  \bibinfo{author}{\bibfnamefont{A.}~\bibnamefont{Losquin}},
  \bibinfo{author}{\bibfnamefont{O.~S.} \bibnamefont{C.~Colliex}},
  \bibinfo{author}{\bibfnamefont{F.~J.} \bibnamefont{{Garc\'{\i}a de Abajo}}},
  \bibnamefont{et~al.}, \bibinfo{journal}{Ultramicroscopy}
  \textbf{\bibinfo{volume}{203}}, \bibinfo{pages}{44} (\bibinfo{year}{2019}).

\bibitem[{\citenamefont{Vanacore et~al.}(2019)\citenamefont{Vanacore, Berruto,
  Madan, Pomarico, Biagioni, Lamb, McGrouther, Reinhardt, Kaminer, Barwick
  et~al.}}]{paper332}
\bibinfo{author}{\bibfnamefont{G.~M.} \bibnamefont{Vanacore}},
  \bibinfo{author}{\bibfnamefont{G.}~\bibnamefont{Berruto}},
  \bibinfo{author}{\bibfnamefont{I.}~\bibnamefont{Madan}},
  \bibinfo{author}{\bibfnamefont{E.}~\bibnamefont{Pomarico}},
  \bibinfo{author}{\bibfnamefont{P.}~\bibnamefont{Biagioni}},
  \bibinfo{author}{\bibfnamefont{R.~J.} \bibnamefont{Lamb}},
  \bibinfo{author}{\bibfnamefont{D.}~\bibnamefont{McGrouther}},
  \bibinfo{author}{\bibfnamefont{O.}~\bibnamefont{Reinhardt}},
  \bibinfo{author}{\bibfnamefont{I.}~\bibnamefont{Kaminer}},
  \bibinfo{author}{\bibfnamefont{B.}~\bibnamefont{Barwick}},
  \bibnamefont{et~al.}, \bibinfo{journal}{Nat.\ Mater.}
  \textbf{\bibinfo{volume}{18}}, \bibinfo{pages}{573} (\bibinfo{year}{2019}).

\bibitem[{\citenamefont{Kfir}(2019)}]{K19}
\bibinfo{author}{\bibfnamefont{O.}~\bibnamefont{Kfir}},
  \bibinfo{journal}{Phys.\ Rev.\ Lett.} \textbf{\bibinfo{volume}{123}},
  \bibinfo{pages}{103602} (\bibinfo{year}{2019}).

\bibitem[{\citenamefont{{Di Giulio} et~al.}(2019)\citenamefont{{Di Giulio},
  Kociak, and {Garc\'{\i}a de Abajo}}}]{paper339}
\bibinfo{author}{\bibfnamefont{V.}~\bibnamefont{{Di Giulio}}},
  \bibinfo{author}{\bibfnamefont{M.}~\bibnamefont{Kociak}}, \bibnamefont{and}
  \bibinfo{author}{\bibfnamefont{F.~J.} \bibnamefont{{Garc\'{\i}a de Abajo}}},
  \bibinfo{journal}{Optica} \textbf{\bibinfo{volume}{6}}, \bibinfo{pages}{1524}
  (\bibinfo{year}{2019}).

\bibitem[{\citenamefont{Talebi}(2020)}]{T20}
\bibinfo{author}{\bibfnamefont{N.}~\bibnamefont{Talebi}},
  \bibinfo{journal}{Phys.\ Rev.\ Lett.} \textbf{\bibinfo{volume}{125}},
  \bibinfo{pages}{080401} (\bibinfo{year}{2020}).

\bibitem[{\citenamefont{Kfir et~al.}(2020)\citenamefont{Kfir,
  Louren\c{c}o-Martins, Storeck, Sivis, Harvey, Kippenberg, Feist, and
  Ropers}}]{KLS20}
\bibinfo{author}{\bibfnamefont{O.}~\bibnamefont{Kfir}},
  \bibinfo{author}{\bibfnamefont{H.}~\bibnamefont{Louren\c{c}o-Martins}},
  \bibinfo{author}{\bibfnamefont{G.}~\bibnamefont{Storeck}},
  \bibinfo{author}{\bibfnamefont{M.}~\bibnamefont{Sivis}},
  \bibinfo{author}{\bibfnamefont{T.~R.} \bibnamefont{Harvey}},
  \bibinfo{author}{\bibfnamefont{T.~J.} \bibnamefont{Kippenberg}},
  \bibinfo{author}{\bibfnamefont{A.}~\bibnamefont{Feist}}, \bibnamefont{and}
  \bibinfo{author}{\bibfnamefont{C.}~\bibnamefont{Ropers}},
  \bibinfo{journal}{Nature} \textbf{\bibinfo{volume}{582}}, \bibinfo{pages}{46}
  (\bibinfo{year}{2020}).

\bibitem[{\citenamefont{Wang et~al.}(2020)\citenamefont{Wang, Dahan, Shentcis,
  Kauffmann, Hayun, Reinhardt, Tsesses, and Kaminer}}]{WDS20}
\bibinfo{author}{\bibfnamefont{K.}~\bibnamefont{Wang}},
  \bibinfo{author}{\bibfnamefont{R.}~\bibnamefont{Dahan}},
  \bibinfo{author}{\bibfnamefont{M.}~\bibnamefont{Shentcis}},
  \bibinfo{author}{\bibfnamefont{Y.}~\bibnamefont{Kauffmann}},
  \bibinfo{author}{\bibfnamefont{A.~B.} \bibnamefont{Hayun}},
  \bibinfo{author}{\bibfnamefont{O.}~\bibnamefont{Reinhardt}},
  \bibinfo{author}{\bibfnamefont{S.}~\bibnamefont{Tsesses}}, \bibnamefont{and}
  \bibinfo{author}{\bibfnamefont{I.}~\bibnamefont{Kaminer}},
  \bibinfo{journal}{Nature} \textbf{\bibinfo{volume}{582}}, \bibinfo{pages}{50}
  (\bibinfo{year}{2020}).

\bibitem[{\citenamefont{{Garc\'{\i}a de Abajo}
  et~al.}(2016)\citenamefont{{Garc\'{\i}a de Abajo}, Barwick, and
  Carbone}}]{paper272}
\bibinfo{author}{\bibfnamefont{F.~J.} \bibnamefont{{Garc\'{\i}a de Abajo}}},
  \bibinfo{author}{\bibfnamefont{B.}~\bibnamefont{Barwick}}, \bibnamefont{and}
  \bibinfo{author}{\bibfnamefont{F.}~\bibnamefont{Carbone}},
  \bibinfo{journal}{Phys.\ Rev.\ B} \textbf{\bibinfo{volume}{94}},
  \bibinfo{pages}{041404(R)} (\bibinfo{year}{2016}).

\bibitem[{\citenamefont{Cai et~al.}(2018)\citenamefont{Cai, Reinhardt, Kaminer,
  and {Garc\'{\i}a de Abajo}}}]{paper312}
\bibinfo{author}{\bibfnamefont{W.}~\bibnamefont{Cai}},
  \bibinfo{author}{\bibfnamefont{O.}~\bibnamefont{Reinhardt}},
  \bibinfo{author}{\bibfnamefont{I.}~\bibnamefont{Kaminer}}, \bibnamefont{and}
  \bibinfo{author}{\bibfnamefont{F.~J.} \bibnamefont{{Garc\'{\i}a de Abajo}}},
  \bibinfo{journal}{Phys.\ Rev.\ B} \textbf{\bibinfo{volume}{98}},
  \bibinfo{pages}{045424} (\bibinfo{year}{2018}).

\bibitem[{\citenamefont{Kone\v{c}n\'{a} and {Garc\'{\i}a de
  Abajo}}(2020)}]{paper351}
\bibinfo{author}{\bibfnamefont{A.}~\bibnamefont{Kone\v{c}n\'{a}}}
  \bibnamefont{and} \bibinfo{author}{\bibfnamefont{F.~J.}
  \bibnamefont{{Garc\'{\i}a de Abajo}}}, \bibinfo{journal}{Phys.\ Rev.\ Lett.}
  \textbf{\bibinfo{volume}{125}}, \bibinfo{pages}{030801}
  (\bibinfo{year}{2020}).

\bibitem[{\citenamefont{Schwartz et~al.}(2019)\citenamefont{Schwartz, Axelrod,
  Campbell, Turnbaugh, Glaeser, and M\"uller}}]{SAC19}
\bibinfo{author}{\bibfnamefont{O.}~\bibnamefont{Schwartz}},
  \bibinfo{author}{\bibfnamefont{J.~J.} \bibnamefont{Axelrod}},
  \bibinfo{author}{\bibfnamefont{S.~L.} \bibnamefont{Campbell}},
  \bibinfo{author}{\bibfnamefont{C.}~\bibnamefont{Turnbaugh}},
  \bibinfo{author}{\bibfnamefont{R.~M.} \bibnamefont{Glaeser}},
  \bibnamefont{and} \bibinfo{author}{\bibfnamefont{H.}~\bibnamefont{M\"uller}},
  \bibinfo{journal}{Nat.\ Methods} \textbf{\bibinfo{volume}{16}},
  \bibinfo{pages}{1016} (\bibinfo{year}{2019}).

\bibitem[{Val()}]{ValerioCompton}
\bibinfo{note}{A. Valerio and F. J. Garc\'{\i}a de Abajo, in preparation.}

\bibitem[{\citenamefont{Allen et~al.}(2001)\citenamefont{Allen, Oxley, and
  Paganin}}]{AOP01}
\bibinfo{author}{\bibfnamefont{L.~J.} \bibnamefont{Allen}},
  \bibinfo{author}{\bibfnamefont{M.~P.} \bibnamefont{Oxley}}, \bibnamefont{and}
  \bibinfo{author}{\bibfnamefont{D.}~\bibnamefont{Paganin}},
  \bibinfo{journal}{Phys.\ Rev.\ Lett.} \textbf{\bibinfo{volume}{87}},
  \bibinfo{pages}{123902} (\bibinfo{year}{2001}).

\bibitem[{\citenamefont{Paganin et~al.}(2018)\citenamefont{Paganin, Petersen,
  and Beltran}}]{PPB18}
\bibinfo{author}{\bibfnamefont{D.~M.} \bibnamefont{Paganin}},
  \bibinfo{author}{\bibfnamefont{T.~C.} \bibnamefont{Petersen}},
  \bibnamefont{and} \bibinfo{author}{\bibfnamefont{M.~A.}
  \bibnamefont{Beltran}}, \bibinfo{journal}{Phys.\ Rev.\ A}
  \textbf{\bibinfo{volume}{97}}, \bibinfo{pages}{023835}
  (\bibinfo{year}{2018}).

\bibitem[{\citenamefont{{Hayes} et~al.}(1980)\citenamefont{{Hayes}, {Jae Lim},
  and {Oppenheim}}}]{HLO1980}
\bibinfo{author}{\bibfnamefont{M.}~\bibnamefont{{Hayes}}},
  \bibinfo{author}{\bibnamefont{{Jae Lim}}}, \bibnamefont{and}
  \bibinfo{author}{\bibfnamefont{A.}~\bibnamefont{{Oppenheim}}},
  \bibinfo{journal}{IEEE Transactions on Acoustics, Speech, and Signal
  Processing} \textbf{\bibinfo{volume}{28}}, \bibinfo{pages}{672}
  (\bibinfo{year}{1980}).

\bibitem[{\citenamefont{Aiger and Talbot}(2010)}]{AT10}
\bibinfo{author}{\bibfnamefont{D.}~\bibnamefont{Aiger}} \bibnamefont{and}
  \bibinfo{author}{\bibfnamefont{H.}~\bibnamefont{Talbot}}, in
  \emph{\bibinfo{booktitle}{2010 IEEE Computer Society Conference on Computer
  Vision and Pattern Recognition}} (\bibinfo{year}{2010}), pp.
  \bibinfo{pages}{295--302}.

\bibitem[{\citenamefont{Spurgeon et~al.}(2020)\citenamefont{Spurgeon, Ophus,
  Jones, Petford-Long, Kalinin, Olszta, Dunin-Borkowski, Salmon, Hattar, Yang
  et~al.}}]{SOJ20}
\bibinfo{author}{\bibfnamefont{S.~R.} \bibnamefont{Spurgeon}},
  \bibinfo{author}{\bibfnamefont{C.}~\bibnamefont{Ophus}},
  \bibinfo{author}{\bibfnamefont{L.}~\bibnamefont{Jones}},
  \bibinfo{author}{\bibfnamefont{A.}~\bibnamefont{Petford-Long}},
  \bibinfo{author}{\bibfnamefont{S.~V.} \bibnamefont{Kalinin}},
  \bibinfo{author}{\bibfnamefont{M.~J.} \bibnamefont{Olszta}},
  \bibinfo{author}{\bibfnamefont{R.~E.} \bibnamefont{Dunin-Borkowski}},
  \bibinfo{author}{\bibfnamefont{N.}~\bibnamefont{Salmon}},
  \bibinfo{author}{\bibfnamefont{K.}~\bibnamefont{Hattar}},
  \bibinfo{author}{\bibfnamefont{W.-C.~D.} \bibnamefont{Yang}},
  \bibnamefont{et~al.}, \bibinfo{journal}{Nat.\ Mater.} pp.
  \bibinfo{pages}{DOI: 10.1038/s41563--020--00833--z} (\bibinfo{year}{2020}).

\bibitem[{\citenamefont{Press et~al.}(1992)\citenamefont{Press, Teukolsky,
  Vetterling, and Flannery}}]{PTV92}
\bibinfo{author}{\bibfnamefont{W.~H.} \bibnamefont{Press}},
  \bibinfo{author}{\bibfnamefont{S.~A.} \bibnamefont{Teukolsky}},
  \bibinfo{author}{\bibfnamefont{W.~T.} \bibnamefont{Vetterling}},
  \bibnamefont{and} \bibinfo{author}{\bibfnamefont{B.~P.}
  \bibnamefont{Flannery}}, \emph{\bibinfo{title}{Numerical Recipes}}
  (\bibinfo{publisher}{Cambridge University Press}, \bibinfo{address}{New
  York}, \bibinfo{year}{1992}).

\end{thebibliography}

\end{document}